\documentclass[aps,prd,tightenlines,10pt,notitlepage,amsmath,amssymb,nofootinbib]{revtex4}
\usepackage[utf8]{inputenc}
\usepackage[T2A]{fontenc}
\usepackage[russian,english]{babel}
\usepackage{graphicx,xcolor}
\usepackage{subcaption}
\usepackage{booktabs}
\definecolor{darkblue}{rgb}{0,0,0.7}
\definecolor{darkred}{rgb}{0.7,0,0}
\definecolor{darkgreen}{rgb}{0,0.75,0}
\definecolor{darkbrawn}{rgb}{0.25,0,0.75}
\usepackage[colorlinks, unicode, citecolor=darkblue, linkcolor=darkred, urlcolor=blue]{hyperref}

\allowdisplaybreaks
\begin{document}

\author{Aleksandr V. Karpenko}
\affiliation{Physics Faculty, Moscow State University,  Moscow 119991 Russia}


\author{Sergey P. Vyatchanin}
\affiliation{Physics Faculty, Moscow State University,  Moscow 119991 Russia}
\affiliation{Faculty of Physics, Branch of M.V. Lomonosov Moscow State University in Baku, 1 Universitet street, Baku, AZ1144 Azerbaijan}

\title{Multimode Quantum Effects of Optical-Axis Misalignment in Gravitational-Wave Interferometers}

\date{ \today}
\begin{abstract}
 We develop a multimode quantum model of optical-axis misalignment in a simplified representation of the differential mode of an Advanced LIGO interferometer. The system is described as two coupled optical resonators corresponding to the signal extraction cavity and the arm cavity. Analytical expressions for the mode-mixing coefficients are derived, establishing a direct connection between geometrical perturbations of optical elements and multimode optical dynamics.
Using the developed formalism, we investigate the influence of optical-axis misalignment on quantum-noise-limited sensitivity and injected squeezed states. We show that misalignment between the signal extraction cavity and the arm cavity leads to significant sensitivity degradation, frequency-dependent rotation of the squeezing ellipse, and a reduction of the observable squeezing in the fundamental spatial mode. At the same time, the injected squeezing is not destroyed but redistributed among coupled spatial modes. The resulting mode coupling generates quantum correlations between the fundamental and higher-order modes and leads to spatial-mode entanglement verified using the positive partial transpose criterion. Our results demonstrate that optical-axis misalignment should be treated as a coherent multimode quantum process rather than as a simple optical loss mechanism.
\end{abstract}

\maketitle

\section{Introduction}

The direct observation of gravitational waves has opened a new era of precision measurements and has established laser interferometers as the most sensitive displacement sensors ever constructed. During the fourth observing run (O4), the global network of gravitational-wave detectors, including Advanced LIGO, Advanced Virgo, and GEO600, achieved unprecedented sensitivity, enabling the detection of increasingly distant and weaker astrophysical sources \cite{25CapotePRD,25VirgoAO}. As classical noise sources have been progressively reduced, quantum noise has become one of the dominant limitations over a broad frequency range.

Quantum noise in gravitational-wave detectors originates from vacuum fluctuations entering the interferometer through the dark port. At high frequencies, these fluctuations manifest themselves as shot noise, whereas at low frequencies they produce radiation-pressure back action on the interferometer mirrors. The interplay between these two contributions leads to the Standard Quantum Limit (SQL), originally formulated by Braginsky and collaborators \cite{Braginsky68,80a1BrThVo,BrKh92}. One of the most successful approaches to surpassing the SQL is the injection of squeezed vacuum states into the interferometer output port. This technique is routinely employed in modern gravitational-wave detectors and has become an essential component of their quantum-noise reduction systems \cite{20BuikemaPRD,Acernese2019}.

The effectiveness of squeezing critically depends on the spatial overlap between the injected squeezed field and the optical modes supported by the interferometer. In an ideal system, the optical field occupies only the fundamental Hermite--Gaussian mode HG$_0$. In practice, however, imperfections of the optical system lead to coupling between the fundamental mode and higher-order spatial modes (HOMs). Such coupling can arise from two distinct mechanisms. The first is mode mismatch, caused by differences in beam parameters such as the waist size, waist position, or wavefront curvature. The second is optical-axis misalignment, resulting from transverse displacements or angular deviations of optical elements. Both mechanisms redistribute optical power among spatial modes and may significantly affect the observable squeezing.

An analytical modal formalism for describing optical-axis misalignment and wavefront distortions in complex resonant interferometers was developed by Hefetz \textit{et al.}~\cite{97HefetzJOSAB}. In this approach, the optical field is expanded in a basis of Hermite--Gaussian modes, while free-space propagation, misaligned mirrors, and distorting optical elements are represented by matrix operators acting in the modal space. The formalism was applied to Fabry--Perot cavities, coupled cavities, and recycled Michelson interferometers and provided a general framework for calculating alignment signals and designing angular-control systems. It also demonstrated explicitly that angular misalignment transfers optical power from the fundamental mode to higher-order transverse modes.

The coupling of Hermite--Gaussian modes caused by beam displacement, tilt, and mode mismatch has subsequently been studied in greater detail. Analytical expressions for the corresponding coupling coefficients were derived within the Hermite--Gaussian formalism and are widely used for modelling realistic optical systems \cite{23TaoPRD,21TaoOL}. These approaches provide an efficient description of spatial-mode mixing and form the basis of modern multimode simulations of gravitational-wave detectors.

The influence of spatial-mode mismatch on squeezed-light injection has also attracted considerable attention. External mode mismatch between the squeezed-light source, the interferometer, and the output mode cleaner was investigated in several studies \cite{17ToyraPRD,17FuldaAO,21McCullerPRD}. These works demonstrated that spatial mismatch can significantly degrade the observable squeezing and therefore reduce detector sensitivity. More recently, internal mode mismatch caused by thermal aberrations inside the interferometer \cite{26KunsArxiv} and hyperloss arising from coherent spatial-mode mixing \cite{26KorobroArXiv} were analyzed. These multimode quantum models showed that mode mismatch cannot generally be treated as an ordinary optical loss. Instead, higher-order modes act as additional quantum degrees of freedom that coherently interact with the fundamental mode. As a consequence, internal mode mismatch may produce a frequency-dependent rotation of the squeezing ellipse and generate nontrivial multimode quantum correlations.

Despite these advances, several important aspects remain insufficiently explored. Existing analytical treatments of optical-axis misalignment provide general modal operators and describe the excitation of higher-order modes in complex interferometers. However, the relation between the geometrical displacements of individual optical elements, the relative displacement and tilt of the eigenmode axes of coupled optical subsystems, and the resulting mode-mixing matrices has not yet been fully developed in a form directly applicable to quantum-noise calculations.

In particular, the influence of optical-axis misalignment on the quantum properties of the output field remains poorly understood. Previous studies have primarily focused on alignment sensing, optical-power coupling, squeezing degradation, and the frequency-dependent rotation of the squeezing ellipse. Considerably less attention has been paid to the redistribution of quantum correlations among spatial modes and to the emergence of multimode entanglement. A unified framework linking geometrical perturbations of optical elements, the resulting subsystem-axis mismatch, mode-mixing coefficients, quantum-noise-limited sensitivity, and multimode quantum correlations is therefore still lacking.

In this work, we develop a multimode quantum model of optical-axis misalignment in a simplified representation of the Advanced LIGO interferometer. The analysis is restricted to the differential optical mode, which is represented as a system of two coupled resonators formed by the signal extraction cavity and the arm cavity. Using a matrix description of Hermite--Gaussian spatial modes, we derive analytical expressions for the mode-mixing coefficients produced by the relative displacement and tilt of the optical axes of these subsystems. The proposed formalism establishes a direct connection between geometrical displacements of individual optical elements and the resulting multimode optical dynamics. For simplicity, we restrict the analysis to misalignments occurring within a single transverse plane. We further assume that the optical axis of the injected squeezed field coincides with that of the readout system, i.e., the optical parametric amplifier (OPA) and the output mode cleaner (OMC) are perfectly aligned. 

As an application of the developed model, we investigate the influence of optical-axis misalignment on quantum-noise-limited sensitivity, squeezing degradation, the frequency-dependent rotation of the squeezing ellipse, and the redistribution of squeezing among spatial modes. We further analyze the multimode quantum correlations generated by mode coupling and demonstrate the emergence of spatial-mode entanglement between the fundamental mode and higher-order spatial modes. The obtained results show that, within the idealized lossless model
considered here, in which finite apertures and other optical losses are
neglected, misalignment-induced mode coupling should be regarded as a
coherent multimode quantum process rather than as a simple optical-loss
mechanism.

\section{Physical model}
\label{model}

\begin{figure}[h]
	\includegraphics[width=0.65\textwidth]{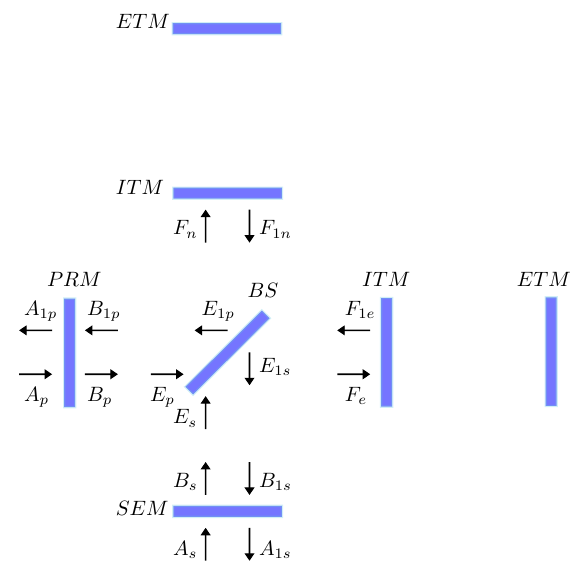}
	\caption{Simplified diagram of a LIGO-type interferometer. The interferometer can be decomposed into
two independent optical systems corresponding to the common (symmetric) and
differential (antisymmetric) optical modes. The common optical mode corresponds to a coupled-cavity system formed by the
power-recycling cavity (PRC) and the arm cavities. The differential optical mode, on the contrary, corresponds to a coupled-cavity
system formed by the signal-extraction cavity (SEC) and the arm cavities. Spatial mode mismatch is assumed to be present only in
the differential optical mode.}
\label{LIGO}
\end{figure}

In this work we consider a simplified multimode model of a LIGO-type
gravitational-wave interferometer. The interferometer can be decomposed into
two independent optical systems corresponding to the common (symmetric) and
differential (antisymmetric) optical modes (see Fig.~\ref{LIGO}). Such a decomposition becomes
possible due to the interferometer tuning in which optical fields entering from
the bright port do not couple to the dark port and vice versa. As a result, the
dynamics of the common and differential optical modes can be considered
independently.

The common optical mode corresponds to a coupled-cavity system formed by the
power-recycling cavity (PRC) \cite{Aasi2015} and the arm cavities. This mode contains the
strong circulating carrier field and is assumed to be perfectly mode matched.
The differential optical mode, on the contrary, corresponds to a coupled-cavity
system formed by the signal-extraction cavity (SEC) \cite{Aasi2015,MiyakawaRSE2006} and the arm cavities. The
injected squeezed vacuum field and the gravitational-wave signal sidebands
propagate in this mode. Spatial mode mismatch is assumed to be present only in
the differential optical mode.

\subsection{Spatial modes and matrix formalism}
\label{spatial_modes}

The optical field inside a resonant optical system is governed by the corresponding wave equation together with the boundary conditions imposed by the optical system. The solutions of this eigenvalue problem form a complete orthonormal set of spatial eigenmodes. Consequently, an arbitrary optical field inside the resonator can be represented as a linear superposition of these modes,

\begin{equation}
E(\mathbf r,t)=\sum_n a_n(t)U_n(\mathbf r),
\label{field_expansion}
\end{equation}
where $U_n(\mathbf r)$ are the resonator eigenmodes describing the spatial structure of the optical field, and $a_n(t)$ are the corresponding complex modal amplitudes describing its temporal evolution. Here and throughout this work, $E(\mathbf r,t)$ denotes the slowly varying complex envelope of the optical field, while the rapidly oscillating optical carrier is omitted.

Instead of describing the complete spatial distribution of the optical field, it is therefore sufficient to specify the vector of modal amplitudes,

\begin{equation}
\mathbf A=
\begin{pmatrix}
a_0\\
a_1\\
a_2\\
\vdots
\end{pmatrix},
\label{mode_vector}
\end{equation}

Within this representation, any linear optical transformation acting on the optical field is completely characterized by its action on the vector of modal amplitudes. Consequently, propagation through an arbitrary linear optical system can be written as

\begin{equation}
\mathbf A'=M\mathbf A,
\label{matrix_form}
\end{equation}

where $M$ is the corresponding transformation matrix. Free-space propagation, reflection and transmission by optical elements, as well as round-trip propagation inside optical resonators, are therefore represented by matrices acting on the modal vector.

As discussed above, the common and differential optical modes of the interferometer are modeled as coupled Fabry--Perot resonators. Consequently, the optical field in each subsystem can be expanded over the eigenmodes of the corresponding resonator. Gravitational-wave interferometers operate with weakly diverging laser beams propagating close to the optical axis. Under these conditions, the paraxial approximation is well satisfied, and the eigenmodes of a Fabry--Perot resonator are given by the Hermite--Gaussian solutions of the paraxial wave equation. Throughout this work, these modes are used as the modal basis.

The two-dimensional Hermite--Gaussian modes are

\begin{align}
U_{nm}(x,y,z)=U_n(x,z)U_m(y,z),
\end{align}

where

\begin{align}
    U_n(x,z)
    &=
    \left(\frac{2}{\pi}\right)^{1/4}
    \left(
        \frac{
            \exp\!\left[i(2n+1)\Psi(z)\right]
        }{
            2^n n! w(z)
        }
    \right)^{1/2}
    H_n\!\left(
        \frac{\sqrt{2}x}{w(z)}
    \right)
    \nonumber\\
    &\times
    \exp\!\left[
        -i\frac{kx^2}{2R_c(z)}
        -
        \frac{x^2}{w^2(z)}
    \right].
\end{align}

Here $H_n$ is the Hermite polynomial, $k$ is the optical wavenumber, and $\Psi(z)$ is the Gouy phase,

\begin{align}
    \Psi(z)
    =
    \arctan\left(
        \frac{z-z_0}{z_R}
    \right),
\end{align}

where

\begin{align}
    z_R
    =
    \frac{\pi w_0^2}{\lambda}
\end{align}

is the Rayleigh range. The beam radius and the wavefront radius of curvature are

\begin{align}
    w(z)
    &=
    w_0
    \sqrt{
        1+
        \left(
            \frac{z-z_0}{z_R}
        \right)^2
    },
    \\
    R_c(z)
    &=
    z-z_0
    +
    \frac{z_R^2}{z-z_0}.
\end{align}

Here $w_0$ is the beam waist radius, $z_0$ is the waist position, and $\lambda$ is the optical wavelength.

The Hermite--Gaussian modes provide an explicit orthonormal basis for representing the spatial structure of the optical field. However, the matrix formalism introduced above does not depend on the particular analytical form of the basis functions. It requires only that the chosen spatial modes constitute a complete orthonormal basis. Throughout the remainder of this work, we therefore denote the basis modes simply by $\{U_n\}$, while the optical field is represented by the corresponding vector of modal amplitudes $A$.

As an example, propagation over a distance $L$ is described by
\begin{equation}
\mathbf A'=M_L\mathbf A,
\label{ML}
\end{equation}
where
\begin{equation}
M_L=
\left\{
\delta_{nm}
e^{-ikL}
e^{i(n+1)\Delta\psi_L}
\right\},
\label{MLmatrix}
\end{equation}

The first exponential factor represents the ordinary propagation phase, whereas the second accounts for the Gouy phase accumulated by each spatial mode. Since $M_L$ is diagonal, different spatial modes propagate independently in the absence of mode mixing.

\subsection{Mode mixing matrices}
\label{mode_mixing}

The diagonal description introduced in the previous section is valid only when
the same spatial-mode basis is used throughout the optical system. In a real
gravitational-wave interferometer, however, different optical subsystems may
possess different eigenmode bases. This situation arises whenever the spatial
distribution of an incoming optical field does not perfectly match the
eigenmodes of the optical system into which it is injected.

As an example, consider an optical field incident on a resonator. If the
spatial distribution of the incident field coincides with the fundamental
eigenmode of the resonator, only the fundamental mode is excited. In contrast,
if the spatial distributions do not coincide, the incident field must be
decomposed over the complete set of resonator eigenmodes. As a result, not only
the fundamental mode but also higher-order spatial modes become excited \cite{Anderson1984}.

 A field that occupies a single spatial mode in one optical subsystem generally becomes
a superposition of several modes when represented in the modal basis of another
subsystem.

Let $\{U_n\}$ denote the spatial modes of one optical subsystem and
$\{V_n\}$ denote the spatial modes of another subsystem. The field expansion
coefficients in these two bases are related through the decomposition
\begin{align}
    U_n
    =
    \sum_s
    \kappa_{sn}
    V_s ,
\end{align}
where $\kappa_{sn}$ are the mode-coupling coefficients.

Physically, these coefficients are overlap integrals between the spatial modes
of the two optical systems \cite{BayerHelms1984},
\begin{align}
    \kappa_{sn}
    = \int V_s^*(\mathbf r) U(\mathbf r) d\mathbf r =
    \langle V_s | U_n \rangle  .
\end{align}

The coefficient $\kappa_{sn}$ determines the fraction of the optical field
initially occupying mode $n$ that is coupled into mode $s$ after the basis
transformation. The larger the overlap between the corresponding spatial modes,
the larger the coupling coefficient.

The mode-coupling coefficients form the mode-mixing matrix
\begin{align}
    \Theta
    =
    \{\kappa_{sn}\}.
\end{align}

The vectors of modal amplitudes in the two bases are therefore related by
\begin{align}
    \mathbf A'
    =
    \Theta \mathbf A.
\end{align}

The matrix $\Theta$ contains the complete information about the spatial mode
mismatch between the two optical subsystems. In the absence of mismatch, the
modal bases coincide and the transformation matrix reduces to the identity
matrix,
\begin{align}
    \Theta
    =
    I.
\end{align}

As the mismatch increases, the off-diagonal elements of $\Theta$ become
nonzero, indicating coupling between different spatial modes.

In the general multimode description introduced in the previous section, the
transformations from one subsystem to another are described by matrix
$\Theta$, while the corresponding transformations in the
opposite direction are described by matrix $U$. The explicit
form of these matrices depends on the physical mechanism responsible for the
mode mismatch.

\subsection{Matrix description of the mismatched interferometer}
\label{sec:matrix_description}

We now apply the modal formalism introduced above to the differential
optical mode of the interferometer. The common optical mode is assumed to
be perfectly mode matched and is used only to establish the stationary
carrier field circulating in the arm cavities. The quantum fluctuations
entering through the antisymmetric port, together with the
gravitational-wave signal sidebands, propagate in the differential mode.

Let $\mathbf A_s$ denote the vector of input spatial-mode amplitudes incident on
the signal extraction mirror, and let $\mathbf B_s$ and $\mathbf B_{1s}$ denote the
intracavity fields propagating away from and toward the signal extraction
mirror, respectively. The boundary condition at the signal extraction
mirror is

\begin{equation}
\mathbf B_s
=
t_s\Theta_1 \mathbf A_s-r_s \mathbf B_{1s},
\label{eq:SEC_boundary}
\end{equation}

where $t_s$ and $r_s$ are the amplitude transmissivity and reflectivity
of the signal extraction mirror. The matrix $\Theta_1$ transforms the
input field from the spatial basis of the injected and readout fields
into the eigenmode basis of the signal extraction cavity.

The round-trip propagation of the differential optical mode through the
coupled SEC--arm-cavity system is described by the matrix

\begin{equation}
S
=
M_{L_s}M_lU_2R\Theta_2M_lM_{L_s}.
\label{eq:round_trip_operator}
\end{equation}

Here $M_{L_s}$ \eqref{eq:app_recycling_propagation} describes propagation from the SEM to the beamsplitter,
$M_l$ \eqref{eq:app_BS_arm_propagation} describes propagation between the beamsplitter and the input test
masses, and $R$ \eqref{eq:app_arm_reflection} is the reflection matrix of the arm cavities. The matrices
$\Theta_2$ and $U_2$ describe the transformations between the eigenmode
bases of the signal extraction cavity and the arm cavities in the forward
and reverse propagation directions, respectively. Thus, $S$ contains
both ordinary modal propagation and coherent spatial-mode mixing within
one complete round trip.

In addition to the purely optical evolution, the differential optical
mode is coupled to the differential displacement of the arm-cavity
mirrors,

\begin{equation}
\Delta x(\Omega)
=
x_n(\Omega)-x_e(\Omega),
\label{eq:differential_displacement}
\end{equation}
where $x_n$ and $x_e$ are the displacements of the north and east arm
mirrors. The field returning to the signal extraction cavity can then be
written in the compact form

\begin{equation}
\mathbf B_{1s}
=
S\mathbf B_s-\mathbf D\,\Delta x,
\label{eq:returning_field_compact}
\end{equation}
where the vector $\mathbf D$ \eqref{eq:app_D_vector} describes the generation and propagation of signal
sidebands produced by the differential mirror displacement. Its explicit
form is given in Appendix~\ref{app:multimode_IO}.

Solving Eqs.~\eqref{eq:SEC_boundary} and
\eqref{eq:returning_field_compact}, the output field at the
antisymmetric port is
\begin{equation}
\mathbf A_{1s}
=
W \mathbf A_s-\mathbf P\,\Delta x,
\label{eq:output_field_compact}
\end{equation}
where
\begin{align}
W
&=
U_1
\left(
I+r_sS
\right)^{-1}
\left(
r_sI+S
\right)
\Theta_1,
\label{eq:W_definition}
\\
\mathbf P
&=
t_sU_1
\left(
I+r_sS
\right)^{-1}
D.
\label{eq:P_definition}
\end{align}
The matrix $W$ describes the purely optical transformation of quantum
fluctuations through the multimode interferometer, whereas $\mathbf P$ describes
the coupling of the differential mirror displacement to the output
field.

The differential displacement obeys the mechanical equation of motion

\begin{equation}
-m\Omega^2\Delta x(\Omega)
=
-mL\Omega^2h(\Omega)
+
F_{\mathrm{BA}}(\Omega),
\label{eq:mechanical_equation_compact}
\end{equation}
where $m$ is the mirror mass, $L$ is the arm-cavity length,
$h(\Omega)$ is the gravitational-wave strain, and
$F_{\mathrm{BA}}$ is the radiation-pressure back-action force generated
by the input quantum fluctuations. In the multimode representation, the
solution can be expressed as

\begin{equation}
\Delta x(\Omega)
=
Lh(\Omega)
-
\mathbf G_q^{T}(\Omega)\mathbf Q_s(\Omega),
\label{eq:mechanical_solution_compact}
\end{equation}
where $\mathbf Q_s$ is the vector of input two-photon quadratures and $\mathbf G_q$ \eqref{eq:app_Gq}
describes the coupling of these quadratures to the radiation-pressure
force.

Following the standard two-photon formalism
\cite{CavesSchumaker1985}, we define the quadrature vector as

\begin{equation}
\mathbf Q_s(\Omega)
=
\left(
a^a_{s0},
a^\phi_{s0},
a^a_{s1},
a^\phi_{s1},
a^a_{s2},
a^\phi_{s2},
\ldots
\right)^T,
\label{eq:quadrature_vector_main}
\end{equation}
with
\begin{align}
a^a_{sn}(\Omega)
&=
\frac{
a_{sn}(\Omega)+a^\dagger_{sn}(-\Omega)
}{
\sqrt{2}
},
\label{eq:amplitude_quadrature_main}
\\
a^\phi_{sn}(\Omega)
&=
\frac{
a_{sn}(\Omega)-a^\dagger_{sn}(-\Omega)
}{
i\sqrt{2}
}.
\label{eq:phase_quadrature_main}
\end{align}

Substitution of Eq.~\eqref{eq:mechanical_solution_compact} into
Eq.~\eqref{eq:output_field_compact} gives the complete multimode
input-output relation

\begin{equation}
\mathbf Q_{1s}(\Omega)
=
Z(\Omega)\mathbf Q_s(\Omega)
-
Lh(\Omega)\mathbf P_q(\Omega),
\label{eq:main_IO_relation}
\end{equation}
where
\begin{equation}
Z(\Omega)
=
W_q(\Omega)
+
\mathbf P_q(\Omega)\mathbf G_q^T(\Omega).
\label{eq:Z_definition_main}
\end{equation}

The first term in Eq.~\eqref{eq:Z_definition_main} describes the purely
optical transformation of the input field, while the second term
represents the radiation-pressure back-action contribution. Explicit
expressions for $W_q$ \eqref{eq:app_Wq}, $\mathbf P_q$ \eqref{eq:app_Pq}, and $\mathbf G_q$ \eqref{eq:app_Gq} are given in
Appendix~\ref{app:multimode_IO}.

The spectral density matrix of the output field is therefore

\begin{equation}
S_1(\Omega)
=
Z(\Omega)S_0(\Omega)Z^\dagger(\Omega),
\label{eq:output_spectral_matrix_main}
\end{equation}
where $S_0$ is the spectral density matrix of the input quadratures. If
only the fundamental spatial mode is phase squeezed, while all
higher-order spatial modes are in vacuum states, then
\begin{equation}
S_0
=
\operatorname{diag}
\left(
e^{2r},
e^{-2r},
1,
1,
\ldots
\right),
\label{eq:input_spectral_matrix_main}
\end{equation}
in the normalization where the vacuum spectral density is equal to unity.

The matrix $S_1(\Omega)$ contains the spectral densities of all output
quadratures and all intermode cross-correlations generated by spatial-mode
mixing and radiation-pressure back action. It therefore provides the
starting point for the analysis of the quantum-noise-limited sensitivity,
the squeezing spectra, the squeezing-angle rotation, and the multimode
quantum correlations presented below. A detailed derivation of
Eqs.~\eqref{eq:returning_field_compact}--\eqref{eq:output_spectral_matrix_main}
is given in Appendix~\ref{app:multimode_IO}.

\subsection{Mode mismatch caused by optical-axis misalignment}
\label{optical_axis_mismatch}

\begin{figure}[h]
	\includegraphics[width=0.35\textwidth]{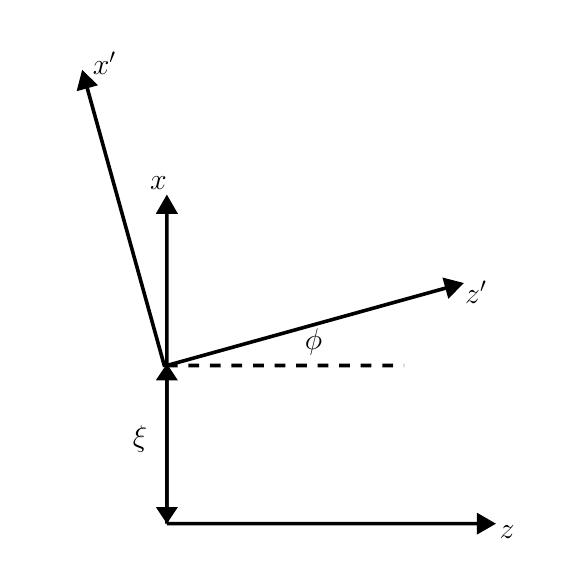}
	\caption{The relative positioning of the basis coordinate systems of various optical subsystems. The z-axes coincide with the optical axes of the subsystems; x represents certain transverse coordinates.  }
\label{xz}
\end{figure}

In this work we focus on spatial mode mismatch caused by a mismatch of the
optical axes of different interferometer subsystems. Such a mismatch may arise
from transverse displacements of optical elements, angular tilts of mirrors, or
a combination of both effects. As a result, the eigenmode basis of one optical
subsystem becomes misaligned with respect to the eigenmode basis of another
subsystem.

Let the optical axis of the incoming beam be displaced by a transverse distance
$\xi$ and tilted by an angle $\phi$ with respect to the optical axis of the
optical system. The corresponding coordinate systems are shown schematically in
Fig.~\ref{xz}.

For paraxial propagation, the coordinates in the two systems are related by
\begin{align}
    z
    &=
    z'\cos\phi
    -
    x'\sin\phi
    \simeq
    z'
    -
    x'\phi,
    \\
    x
    &=
    \xi
    +
    x'\cos\phi
    +
    z'\sin\phi
    \simeq
    \xi
    +
    x'
    +
    z'\phi,
\end{align}
where the small-angle approximation ($\phi\ll 1$) has been used.

The spatial distribution of the optical field expressed in the coordinates of
the second optical system can therefore be written as
\begin{align}
    U_n(x)
    \rightarrow
    U_n(\xi+x')
    e^{ik\phi x'}.
\end{align}

The first factor describes the transverse displacement of the beam, whereas
the second factor represents the linear phase gradient associated with the
angular tilt.

Using the translation operator, this transformation can be rewritten as
\begin{align}
    U_n(x)
    \rightarrow
    e^{ik\phi x'}
    e^{\xi \frac{d}{dx'}}
    U_n(x) .
\end{align}

The  mode-coupling coefficients 
\begin{align}
    \kappa_{sn}
    &=
    \langle U_s|
    e^{ik\phi x'}
    e^{\xi \frac{d}{dx'}}
    |U_n\rangle .
\end{align}

Applying the Baker--Campbell--Hausdorff formula yields
\begin{align}
    \kappa_{sn}
    &=
    e^{-ik\phi\xi/2}
    \langle U_s|
    e^{ik\phi x'
    +
    \xi \frac{d}{dx'}}
    |U_n\rangle .
\end{align}

To evaluate these coefficients, it is convenient to represent the spatial modes
using the harmonic-oscillator basis
\begin{align}
    |U_n\rangle
    &=
    e^{i(2n+1)\Psi/2}
    e^{-ikx^2/(2R_c)}
    |n\rangle ,
\end{align}
where $\Psi$ is the Gouy phase and $R_c$ is the wavefront radius of
curvature.

Substituting this expression into the overlap integrals gives
\begin{align}
    \kappa_{sn}
    &=
    e^{-ik\phi\xi/2}
    e^{i(n-s)\Psi}
    \langle s|
    e^{\xi\frac{d}{dx}
    +
    ik\left(
        \phi-\frac{\xi}{R_c}
    \right)x}
    |n\rangle .
\end{align}

Introducing the annihilation and creation operators
\begin{align}
    \frac{d}{dx}
    &=
    \frac{\hat a-\hat a^\dagger}{w}, \quad
    x
    =
    \frac{w}{2}
    \left(
        \hat a+\hat a^\dagger
    \right),
\end{align}
the coupling coefficients can be written as
\begin{align}
\label{kappa}
    \kappa_{sn}
    &=
    e^{-ik\phi\xi/2}
    e^{i(n-s)\Psi}
    \langle s|
    e^{C\hat a^\dagger-C^*\hat a}
    |n\rangle ,
\end{align}
where
\begin{align}
\label{c}
    C
    &=
    -\frac{\xi}{w}
    +
    i\frac{kw}{2}
    \left(
        \phi-\frac{\xi}{R_c}
    \right).
\end{align}

The matrix elements $\kappa_{sn}$ form the mode-mixing matrix
\begin{align}
    \Theta
    =
    \{\kappa_{sn}\}.
\end{align}

The mode-mixing matrix $\Theta$ obtained here is the well-known Bayer–Helms coupling matrix \cite{BayerHelms1984} describing Hermite-Gaussian mode coupling produced by beam displacement, tilt, and mode mismatch.

The complex mismatch parameter $C$ completely characterizes the optical-axis
misalignment. Its real part describes mode coupling induced by transverse beam
displacement, whereas its imaginary part corresponds to coupling produced by
angular misalignment and wavefront-curvature mismatch.

\begin{figure}[t]
\centering

\begin{subfigure}{0.32\textwidth}
    \centering
    \includegraphics[width=\linewidth]{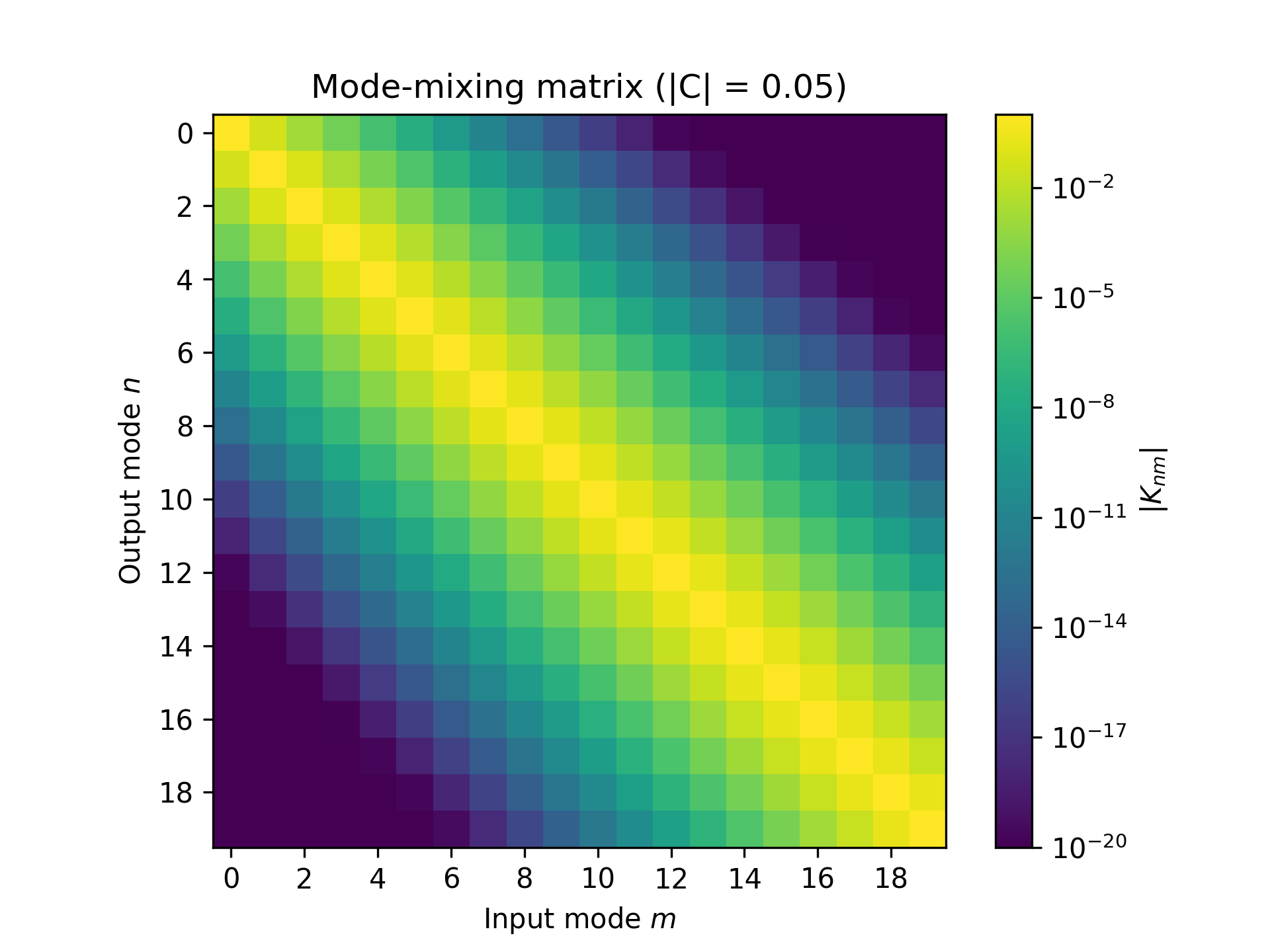}
\end{subfigure}
\hfill
\begin{subfigure}{0.32\textwidth}
    \centering
    \includegraphics[width=\linewidth]{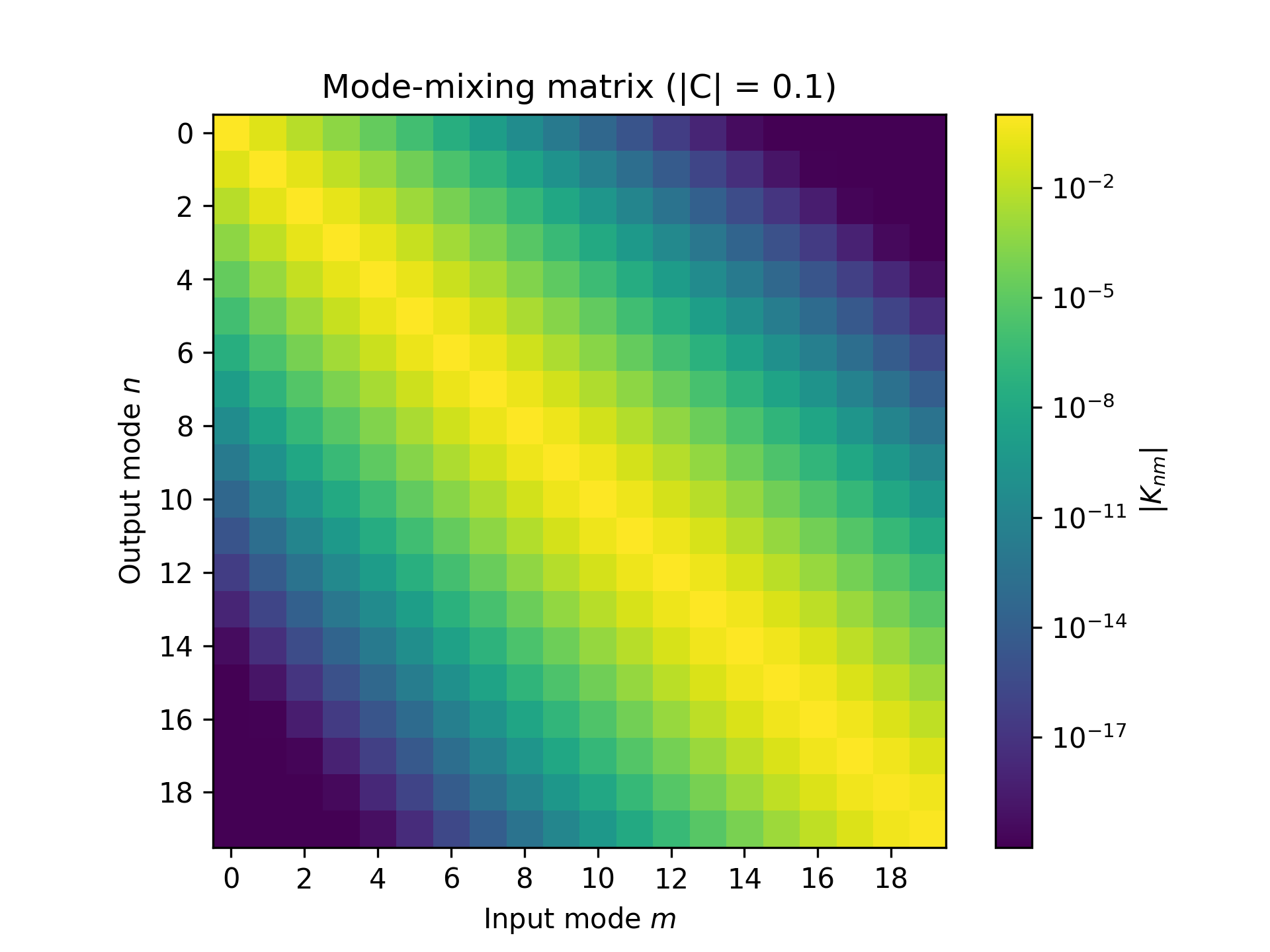}
\end{subfigure}
\hfill
\begin{subfigure}{0.32\textwidth}
    \centering
    \includegraphics[width=\linewidth]{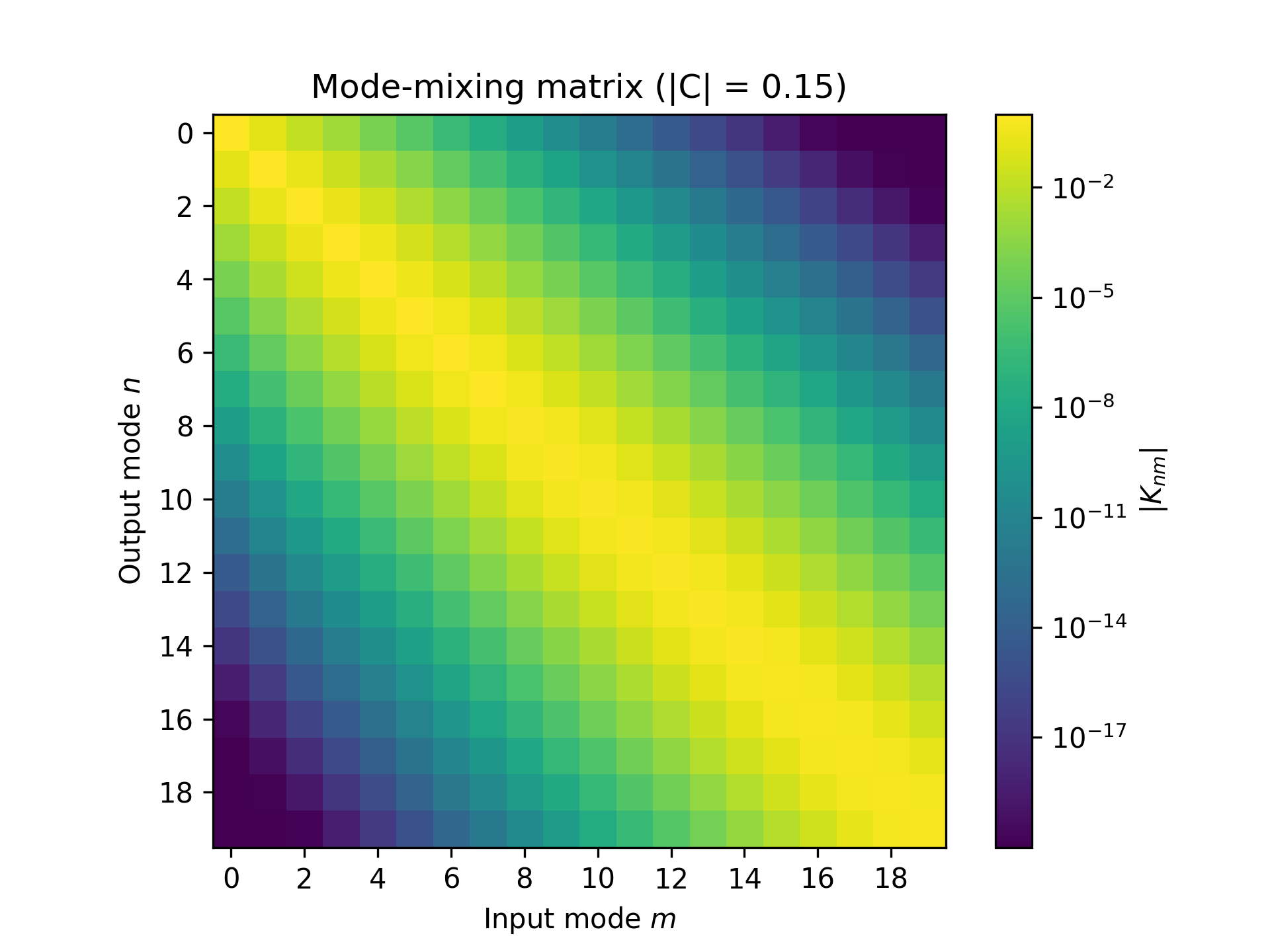}
\end{subfigure}

\caption{
Absolute values of the mode-mixing matrix elements $|\kappa_{nm}|$ describing coupling between Hermite--Gaussian spatial modes caused by optical-axis misalignment. The horizontal axis corresponds to the input mode index $m$, while the vertical axis corresponds to the output mode index $n$. The three panels show the mode-mixing matrices calculated for increasing values of the misalignment parameter $|C|$. As the misalignment increases, the coupling spreads over a wider range of spatial modes, indicating stronger redistribution of optical power and quantum correlations between higher-order modes.
}
\label{MixingMatrix}
\end{figure}

Figure~\ref{MixingMatrix} shows the absolute values of the mode-mixing matrix elements $|\kappa_{nm}|$ calculated for several values of the misalignment parameter $|C|$. The diagonal structure of the matrices indicates that the strongest coupling occurs between spatial modes with similar mode indices. As the misalignment increases, the width of the coupling region grows, and an increasing number of higher-order modes become involved in the optical dynamics.

This behavior reflects the physical nature of optical-axis misalignment. A displaced optical axis does not generate an arbitrary redistribution of optical fields but rather produces a gradual transfer of optical power from a given mode into neighboring Hermite--Gaussian modes. Consequently, larger values of $|C|$ lead to stronger multimode coupling and more pronounced redistribution of squeezed states among higher-order spatial modes. As will be shown in the following sections, this redistribution is responsible for both the degradation of the observed squeezing in the fundamental mode and the appearance of multimode quantum correlations.

For the reverse transformation, corresponding to propagation from the optical
system back to the input-output beam basis, direct calculation yields
\begin{align}
    U
    =
    \Theta^T .
\end{align}

Therefore, the entire multimode dynamics of the mismatched interferometer can
be fully expressed in terms of the mode-mixing matrix $\Theta$ and the
mismatch parameter $C$ determining its elements.

\subsection{Differential mode with optical-axis mismatch}
\begin{figure}[h]
\includegraphics[width=0.8\textwidth]{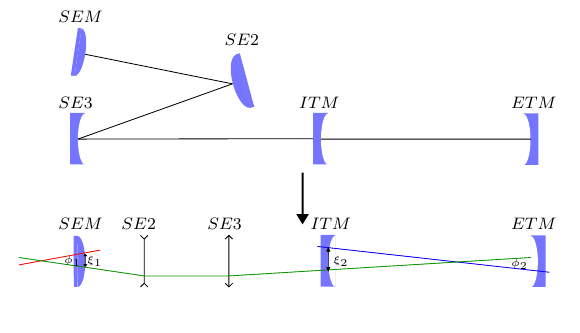}\\
    \caption{Differential optical mode of a signal-recycled interferometer and its simplified representation used in this work. The differential mode consists of the signal extraction cavity (SEC), formed by the mirrors SEM, SE2, and SE3, coupled to the arm cavity formed by the ITM and ETM mirrors. In the simplified model, the folding mirrors SE2 and SE3 are replaced by equivalent thin lenses, which preserve the optical-axis geometry of the SEC while simplifying the analytical description of mode coupling. The lower panel illustrates the optical axes of the three subsystems considered in the model: the readout system (red), the signal extraction cavity (green), and the arm cavity (blue). The relative transverse displacements $\xi_1$, $\xi_2$ and angular deviations $\phi_1$, $\phi_2$ between these optical axes give rise to the mode-mixing coefficients $C_1$ and $C_2$, which characterize coupling between Hermite--Gaussian spatial modes.}\label{SEC_scheme}
\end{figure}
The signal extraction cavity (SEC) of a gravitational-wave interferometer is not a simple two-mirror resonator. As shown schematically in Fig.~\ref{SEC_scheme}, it is a folded optical cavity formed by the signal extraction mirror (SEM), the folding mirrors SR$_2$ and SR$_3$, and the input test masses (ITM). The optical field propagates through the cavity along a folded trajectory and is reflected from the mirrors SR$_2$ and SR$_3$ at nonzero incidence angles.

For the analysis of Gaussian beam propagation, it is convenient to replace this folded geometry with an equivalent one-dimensional optical system. Within the paraxial approximation, the action of the folding mirrors can be represented by equivalent thin lenses, while the remaining mirrors are treated as spherical reflecting surfaces. As a result, the SEC can be modeled as an equivalent linear resonator consisting of spherical mirrors and thin lenses. Such a representation preserves the Gaussian eigenmode of the cavity and allows the use of the standard ABCD-matrix formalism \cite{KogelnikLi1966} for calculating both the cavity eigenmode and its optical axis.

The optical axis of a resonator is defined as the trajectory along which a paraxial ray propagates through the optical system and reproduces itself after a complete round trip. In an ideally aligned interferometer, the optical axes of all coupled optical subsystems coincide. Consequently, the spatial eigenmodes of the signal extraction cavity, the arm cavities, and the input optical system are identical, and no coupling between different spatial modes occurs.

In a realistic interferometer, optical elements are subject to small transverse displacements arising from fabrication tolerances, thermal distortions, seismic motion, and alignment imperfections. Since these displacements are generally independent and uncorrelated, the optical axes of different subsystems become displaced with respect to one another. As a consequence, the spatial eigenmodes of the coupled cavities no longer coincide, leading to mode mismatch and coupling between the fundamental and higher-order spatial modes.

In the following analysis, all transverse displacements are assumed to occur in a single plane $OZX$. Under this assumption, the problem becomes effectively one-dimensional and can be described by a single transverse coordinate $x$. The displacement of each optical element results in a corresponding displacement and tilt of the optical axis. These quantities can be calculated using the ABCD-matrix formalism for misaligned optical systems \cite{TovarCasperson1995}. This formalism  yields the displacement and tilt of its optical axis as functions of mirror and lens displacements. Therefore, the mode-mismatch parameters $C$ \eqref{c} can ultimately be expressed in terms of physical displacements of the optical components.

In the model developed in this work, transitions between different spatial-mode bases occur only at two locations. The first transition takes place at the signal extraction mirror (SEM), where the spatial basis of the incoming optical field is transformed into the eigenmode basis of the signal extraction cavity. The corresponding mode-mixing matrix is characterized by the mismatch parameter $C_1$.

The second transition occurs at the input test mass (ITM), where the spatial basis of the signal extraction cavity is transformed into the spatial basis of the arm cavity. The corresponding mode-mixing matrix is characterized by the mismatch parameter $C_2$.

The parameter $C_1$ is generally complex. Physically, this reflects the fact that the optical axis of the incoming beam, which in the present model coincides with the optical axis of the readout system, may possess both an arbitrary transverse displacement and an arbitrary angular tilt with respect to the optical axis of the signal extraction cavity. Consequently, the mismatch contains independent displacement-like and tilt-like contributions.

The situation is considerably simpler for the parameter $C_2$. In the present model all cavity mirrors are assumed to be spherical. The optical axis of a stable resonator must intersect a mirror surface along the local surface normal. Since the normal to a spherical surface always passes through its center of curvature, both the optical axis of the signal extraction cavity and the optical axis of the arm cavity pass through the center of curvature of the ITM.

Consequently, the two optical axes always intersect at this point and are therefore not independent. A transverse displacement of the beam spot on the ITM uniquely determines the relative angular tilt between the two axes. The relative mismatch between the cavities can therefore be completely characterized by a single parameter describing the transverse displacement of the optical axes at the mirror surface.

The angular difference $\phi$ between the axes is uniquely related to this displacement $\xi$ through the mirror curvature $R_c$:

\begin{equation}
\phi = \frac{\xi}{R_c},
\end{equation}

 The displacement and angular contributions are therefore not independent, and the corresponding mismatch parameter $C_2$ becomes purely real (see Eq.\eqref{c}).

As shown in the previous sections, the complete multimode dynamics of the mismatched interferometer can be fully expressed through the mode-mixing matrices. Consequently, once the optical-axis displacements and tilts are related to the physical displacements of mirrors and lenses using the ABCD formalism, the entire problem reduces to determining the mismatch parameters $C_1$ and $C_2$. These parameters define the transformations between the spatial-mode bases of the coupled optical subsystems and therefore completely characterize the influence of optical-axis mismatch on the quantum-noise properties of the interferometer.

\section{Results}

In this section, we investigate the impact of optical mode mismatch on the
quantum-noise-limited sensitivity of the interferometer. To isolate the effects
associated with spatial mode coupling, only quantum noise is taken into account,
whereas all classical noise sources, including thermal, seismic, suspension,
and technical noises, are neglected. Consequently, the sensitivity curves
presented below should be interpreted as quantum-noise-limited sensitivities. All optical elements are assumed to have infinite transverse apertures.
Consequently, diffraction and clipping losses of higher-order spatial modes
are neglected. No other optical losses are included in the present model.

Throughout this section, a frequency independent phase-squeezed vacuum state with a squeezing level
of $12\,\mathrm{dB}$ is injected into the antisymmetric port of the
interferometer. The squeezed field is assumed to occupy only the fundamental
spatial mode, while all higher-order spatial modes are in vacuum states. The
readout quadrature is fixed to the phase quadrature, and no frequency-dependent
optimization of the squeezing angle is applied. The OPA and OMC are assumed to be perfectly aligned, so that no additional
external misalignment is introduced between the injected and detected
spatial-mode bases.

All numerical calculations presented in this work were performed using the interferometer parameters listed in Table~\ref{parameters}. Unless otherwise stated, the optical field was expanded over the first N=20 Hermite--Gaussian spatial modes. This truncation was found to provide converged results for all quantities presented below.

\begin{table}[t]
\centering
\caption{Parameters used in the numerical simulations.}
\label{parameters}
\begin{tabular}{lll}
\toprule
Parameter & Symbol & Value \\
\midrule
Arm cavity length & $L_{\rm arm}$ & $3995~\mathrm{m}$ \\
Signal extraction cavity length & $L_s$ & $55~\mathrm{m}$ \\
One-way SEC Gouy phase & $\Delta\psi_{L_s}$ & $20~\mathrm{deg}$\\
SEC phase offset & $\delta_{SEC}$ & 0 rad\\
Mirror mass & $m$ & $39.6~\mathrm{kg}$ \\
Laser wavelength & $\lambda$ & $1064~\mathrm{nm}$ \\
Circulating arm power & $P_{\rm arm}$ & $750.6~\mathrm{kW}$ \\
ITM power transmissivity & $T_{\rm ITM}$ & $0.014$ \\
SEC mirror transmissivity & $T_s$ & $0.325$ \\
Power recycling mirror transmissivity & $T_p$ & $0.03$ \\
Injected squeezing & $r$ & $12~\mathrm{dB}$ \\
Number of spatial modes & $N$ & $20$ \\
\bottomrule
\end{tabular}
\end{table}

\subsection{Quantum-noise-limited sensitivity degradation due to mode mismatch}
\label{qn_sensitivity_mismatch}

\begin{figure}[t]
\centering

\begin{subfigure}{0.48\textwidth}
    \centering
    \includegraphics[width=\linewidth]{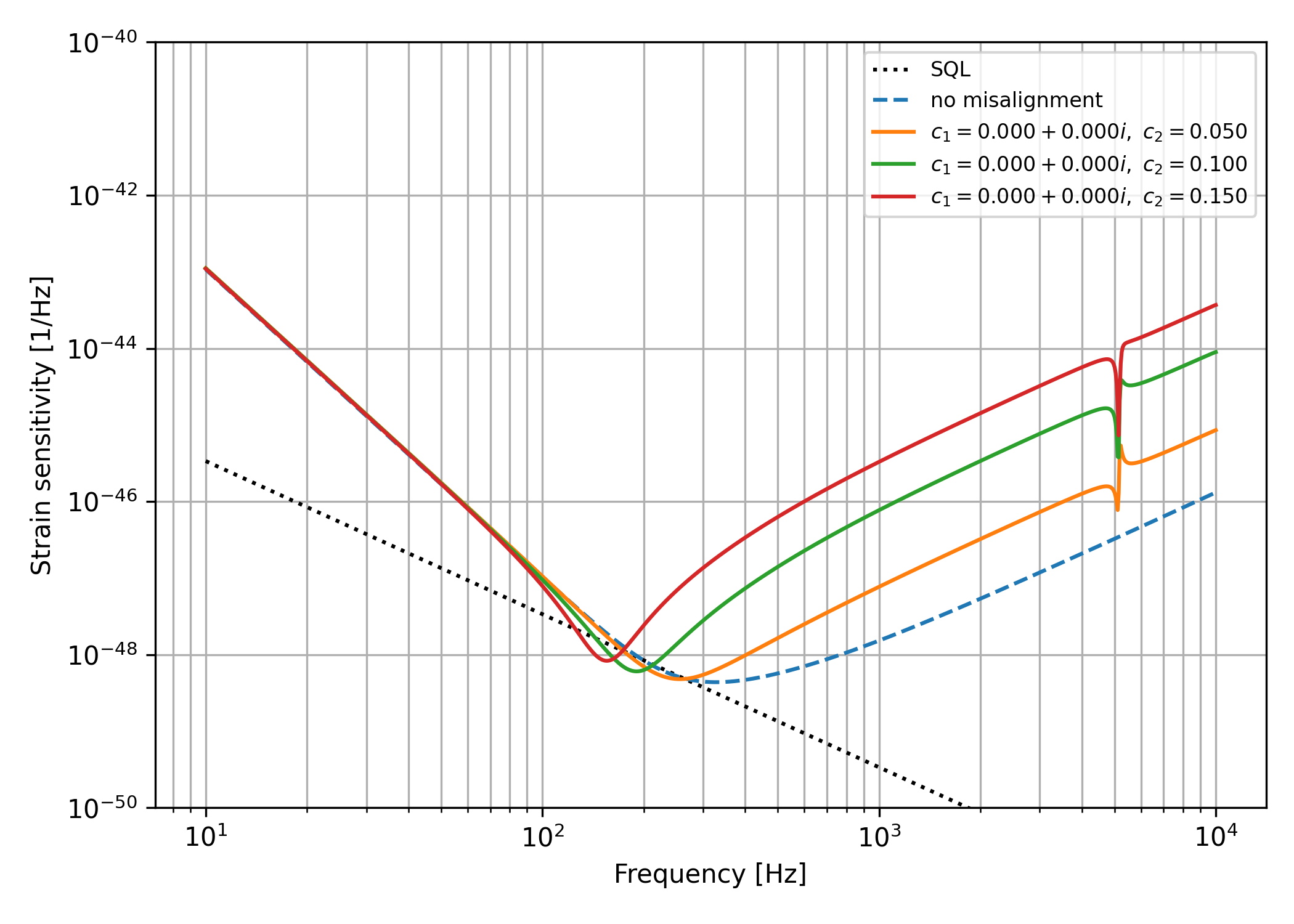}
    \caption{$C_1=0$, $C_2\neq0$.}
    \label{sens_c2_only}
\end{subfigure}
\hfill
\begin{subfigure}{0.48\textwidth}
    \centering
    \includegraphics[width=\linewidth]{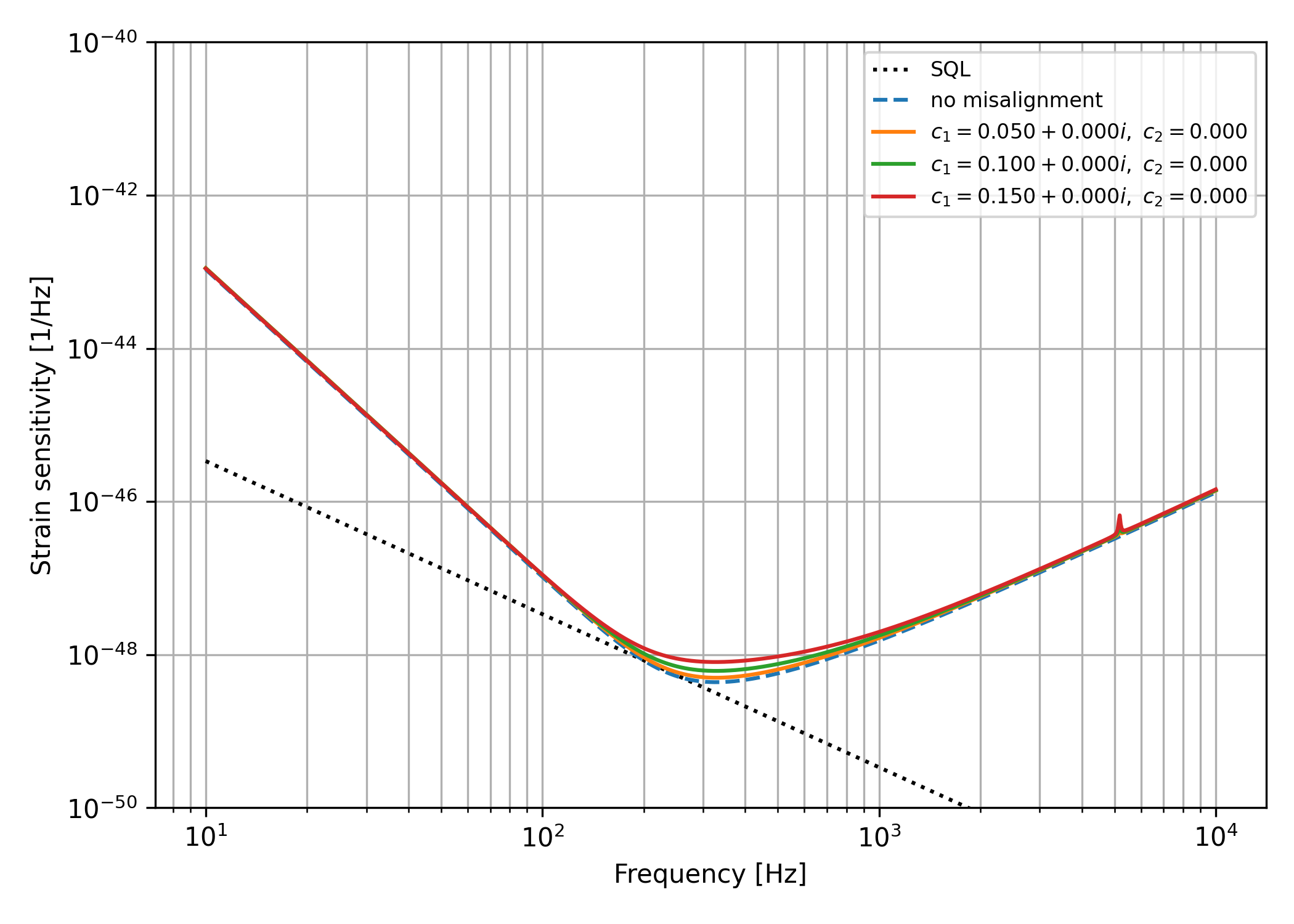}
    \caption{Real $C_1\neq0$, $C_2=0$.}
    \label{sens_c1_real}
\end{subfigure}

\vspace{0.3cm}

\begin{subfigure}{0.48\textwidth}
    \centering
    \includegraphics[width=\linewidth]{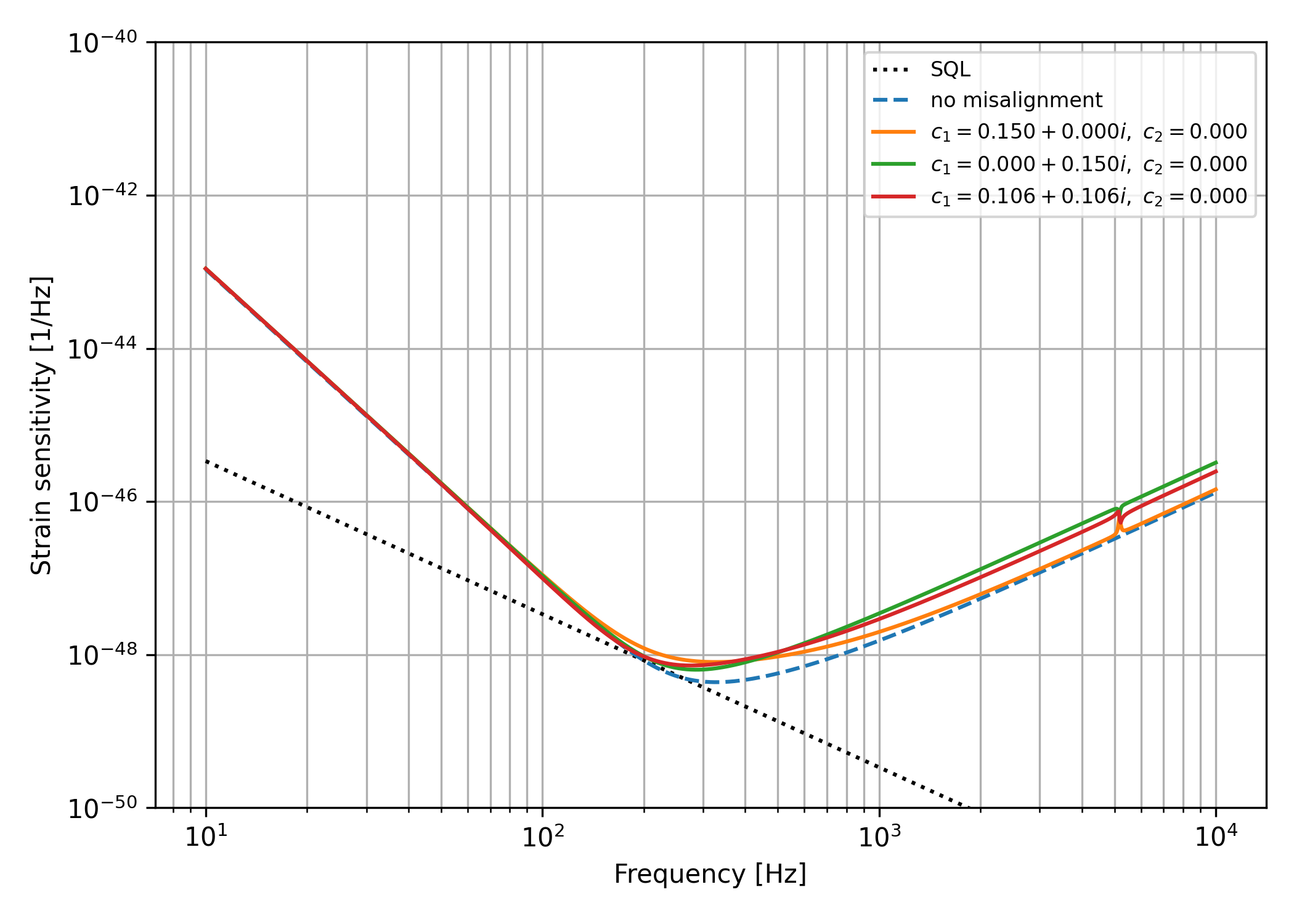}
    \caption{Complex $C_1\neq0$, $C_2=0$.}
    \label{sens_c1_complex}
\end{subfigure}
\hfill
\begin{subfigure}{0.48\textwidth}
    \centering
    \includegraphics[width=\linewidth]{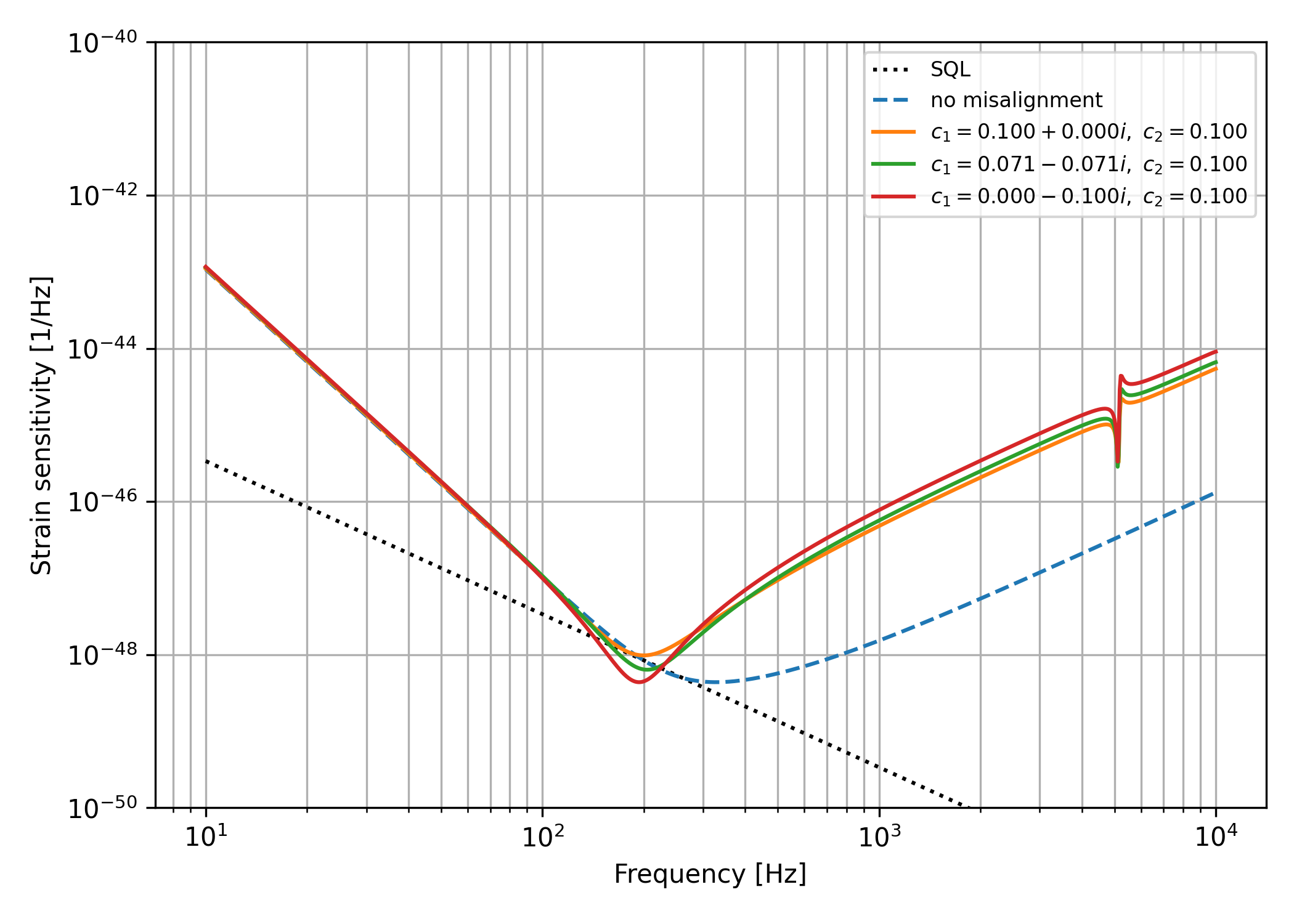}
    \caption{$C_1\neq0$, $C_2\neq0$.}
    \label{sens_c1_c2}
\end{subfigure}

\caption{
Quantum-noise-limited strain sensitivity calculated using Eq.\eqref{eq:main_IO_relation} for
different mode-mismatch configurations.
Panel~(a) shows the case where only the mismatch between the signal extraction
cavity and the arm cavity is present, $C_1=0$ and $C_2\neq0$.
Panel~(b) corresponds to a purely real mismatch between the readout system and
the signal extraction cavity, $C_1\neq0$ and $C_2=0$.
Panel~(c) shows the effect of a complex $C_1$ with $C_2=0$.
Panel~(d) corresponds to the general case where both mismatch parameters are
nonzero.
The blue dashed curve denotes the perfectly matched interferometer, while the
black dotted line indicates the Standard Quantum Limit (SQL).
}
\label{strain_sensitivity_mismatch}
\end{figure}

The quantum-noise-limited strain sensitivity curves shown in
Fig.~\ref{strain_sensitivity_mismatch} were calculated using Eq.\eqref{eq:main_IO_relation} for
different values of the mismatch parameters $C_1$ and $C_2$. The dashed blue
curve corresponds to the perfectly matched interferometer, while the colored
curves represent different levels of mode mismatch. The black dotted line
indicates the Standard Quantum Limit (SQL).

The results demonstrate that the two mismatch mechanisms considered in this
work affect the detector sensitivity in qualitatively different ways. The
parameter $C_1$, which describes the mismatch between the readout system and
the signal extraction cavity, produces only a relatively weak degradation of
sensitivity. Even for comparatively large values of $C_1$, the resulting
sensitivity curves remain close to the perfectly matched case over the entire
detection band. In addition, the sensitivity is found to depend only weakly on
the phase of $C_1$. For a fixed value of $|C_1|$, changing the phase of the
coupling coefficient leads to only minor modifications of the sensitivity
curves. Therefore, the overall sensitivity degradation is governed mainly by
the strength of the mode coupling rather than by its phase.

In contrast, the parameter $C_2$, which characterizes the mismatch between the
eigenmodes of the signal extraction cavity and those of the arm cavities, has a
significantly stronger impact. As $C_2$ increases, the high-frequency
sensitivity deteriorates rapidly, and pronounced modifications of the
quantum-noise spectrum become visible. For sufficiently large mismatch values,
the sensitivity degradation reaches several orders of magnitude at high
frequencies.

At the same time, for some values of $C_2$, the sensitivity becomes slightly
better than that of the perfectly matched interferometer within a limited
frequency range around a few hundred hertz, approaching or even crossing the
SQL. As will be shown in the following section, this behavior is associated
with a mismatch-induced rotation of the squeezing ellipse. In the present
model, a frequency-independent squeezed state is injected into the
interferometer, while the detected quadrature is not optimal at all
frequencies. Consequently, the additional rotation of the squeezing ellipse
caused by mode mismatch may accidentally move the measured quadrature closer
to the optimal one. The observed sensitivity improvement is therefore a
consequence of this accidental quadrature rotation rather than a genuine
reduction of quantum noise.

It should be emphasized that this effect does not imply that mode mismatch is
beneficial. Modern gravitational-wave detectors employ filter cavities to
generate nearly optimal frequency-dependent squeezing \cite{Kimble2001}. In such systems, the
squeezing angle is already close to the optimum value required for broadband
quantum-noise reduction. Any additional mismatch-induced rotation would
generally move the squeezing ellipse away from this optimum and therefore
degrade the detector sensitivity. Consequently, mode mismatch should be
regarded as an undesirable effect that limits the achievable quantum-noise
suppression.

In addition to the overall degradation of sensitivity, pronounced spectral features appear in the vicinity of $5\,\mathrm{kHz}$. These structures originate from the resonant response of the first higher-order spatial mode HG$_1$, which becomes coupled to the fundamental mode due to optical-axis mismatch. In the absence of mode mismatch, the injected squeezed field occupies only the fundamental spatial mode, and the HG$_1$ resonance remains decoupled from the readout channel. However, the mode-mixing process described in the Section \ref{model} transfers a fraction of the optical field into higher-order modes, making their resonant response observable in the detected quantum noise. Depending on the relative amplitudes and phases of the interfering modal contributions, the HG$_1$ resonance may manifest itself either as a peak or as a dip in the sensitivity curve. The position of these spectral features is therefore determined by the resonance frequency of the HG$_1$ mode, while their magnitude increases with increasing mode mismatch.

The observed modifications of the sensitivity curves indicate that mode
mismatch does not simply introduce optical losses. Instead, it generates
coherent coupling between the fundamental spatial mode and higher-order modes,
leading to a redistribution of quantum correlations within the multimode
optical field. To understand the physical origin of this behavior, we next
analyze how mode mismatch modifies the squeezing properties of the fundamental
mode and how quantum correlations emerge between spatial modes.

\section{Discussion}

\subsection{Degradation and rotation of the fundamental-mode squeezing}
\label{fundamental_mode_squeezing}

The results presented in the previous subsection \ref{qn_sensitivity_mismatch} demonstrate that mode mismatch can
substantially modify the quantum-noise-limited sensitivity of the interferometer.
To understand the physical origin of these changes, we analyze how mode mismatch
affects the quantum state of the fundamental spatial mode.

In the absence of mode mismatch, only the fundamental spatial mode contains the
injected squeezed vacuum field, while all higher-order spatial modes remain in
vacuum states. The detected quantum noise is therefore determined by the
squeezing properties of the fundamental mode. When mode mismatch is introduced,
coherent coupling between spatial modes redistributes quantum fluctuations among
different transverse modes. As a result, the squeezing properties of the
fundamental mode are modified.

For each frequency, we extract from the full spectral density matrix $S_1$ \eqref{eq:output_spectral_matrix_main} the
$2\times2$ block corresponding to the $n$-th spatial mode,
\begin{equation}
S_n(\Omega)
=
\begin{pmatrix}
S_{a^a_n a^a_n}(\Omega) & S_{a^a_n a^{\phi}_n}(\Omega) \\
S_{a^{\phi}_n a^a_n}(\Omega) & S_{a^{\phi}_n a^{\phi}_n}(\Omega)
\end{pmatrix}.
\end{equation}

The spectral density of an arbitrary measurable quadrature of this mode,
\begin{equation}
q_{n,\theta}
=
a^{a}_n \cos\theta
+
a^{\phi}_n \sin\theta ,
\end{equation}
is given by
\begin{equation}
S_{n,\theta}(\Omega)
=
\mathbf e_\theta^{T}
\,\mathrm{Re}\!\left[S_n(\Omega)\right]\,
\mathbf e_\theta ,
\qquad
\mathbf e_\theta
=
\begin{pmatrix}
\cos\theta \\
\sin\theta
\end{pmatrix}.
\end{equation}

The squeezed and anti-squeezed quadrature spectral densities are then obtained
as the minimum and maximum eigenvalues of the real symmetric matrix
$\mathrm{Re}\!\left[S_n(\Omega)\right]$:
\begin{align}
S_n^{\min}(\Omega)
&=
\lambda_{\min}
\left[
\mathrm{Re}\,S_n(\Omega)
\right],
\\
S_n^{\max}(\Omega)
&=
\lambda_{\max}
\left[
\mathrm{Re}\,S_n(\Omega)
\right].
\end{align}

The corresponding eigenvectors determine the orientation of the noise ellipse
and define the squeezing angle $\theta_n(\Omega)$. For the fundamental mode,
this angle is denoted by $\theta_0(\Omega)$.

\begin{figure}[t]
\centering

\begin{subfigure}{0.48\textwidth}
    \centering
    \includegraphics[width=\linewidth]{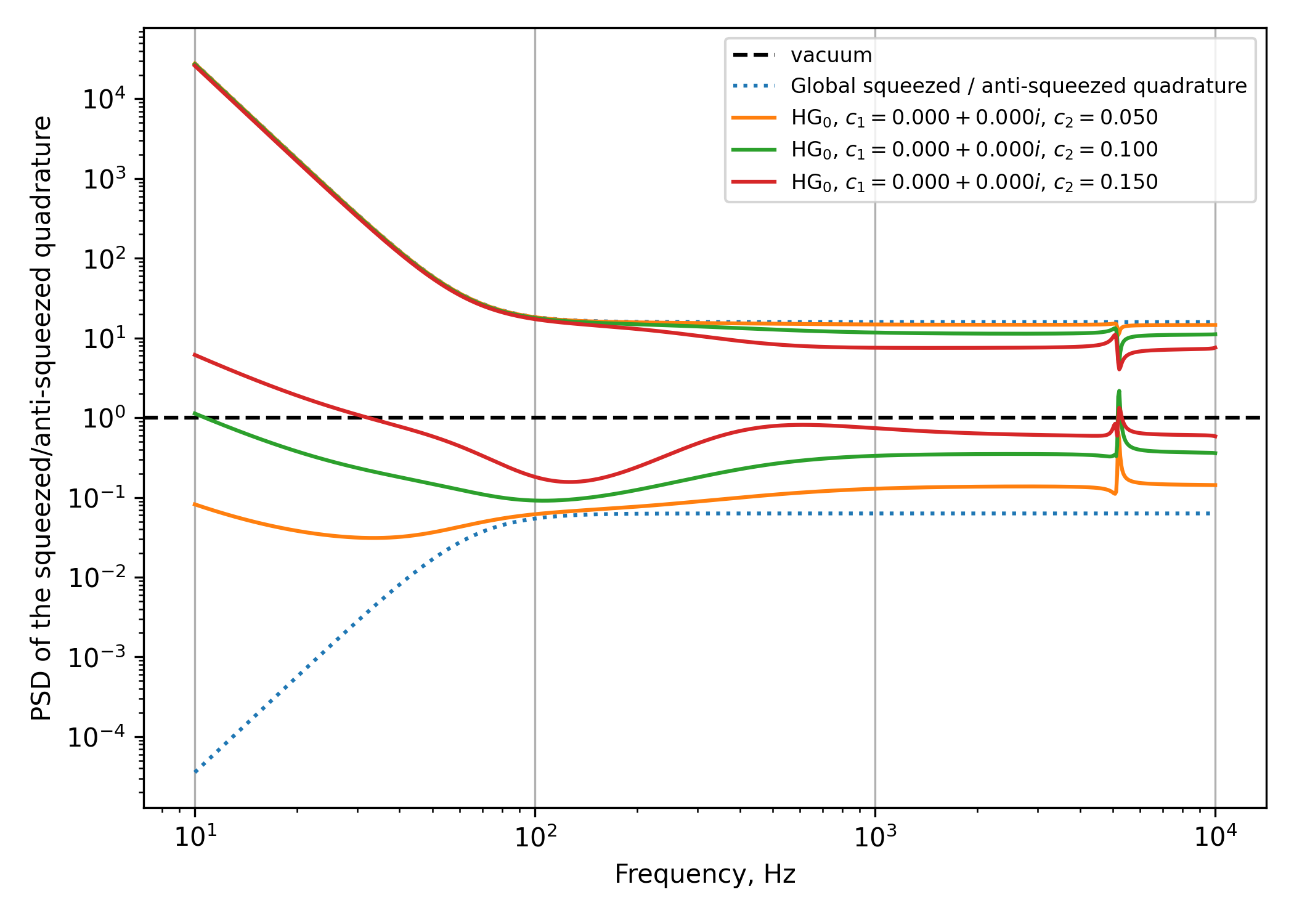}
    \caption{$C_1=0$, $C_2\neq0$.}
    \label{sq1}
\end{subfigure}
\hfill
\begin{subfigure}{0.48\textwidth}
    \centering
    \includegraphics[width=\linewidth]{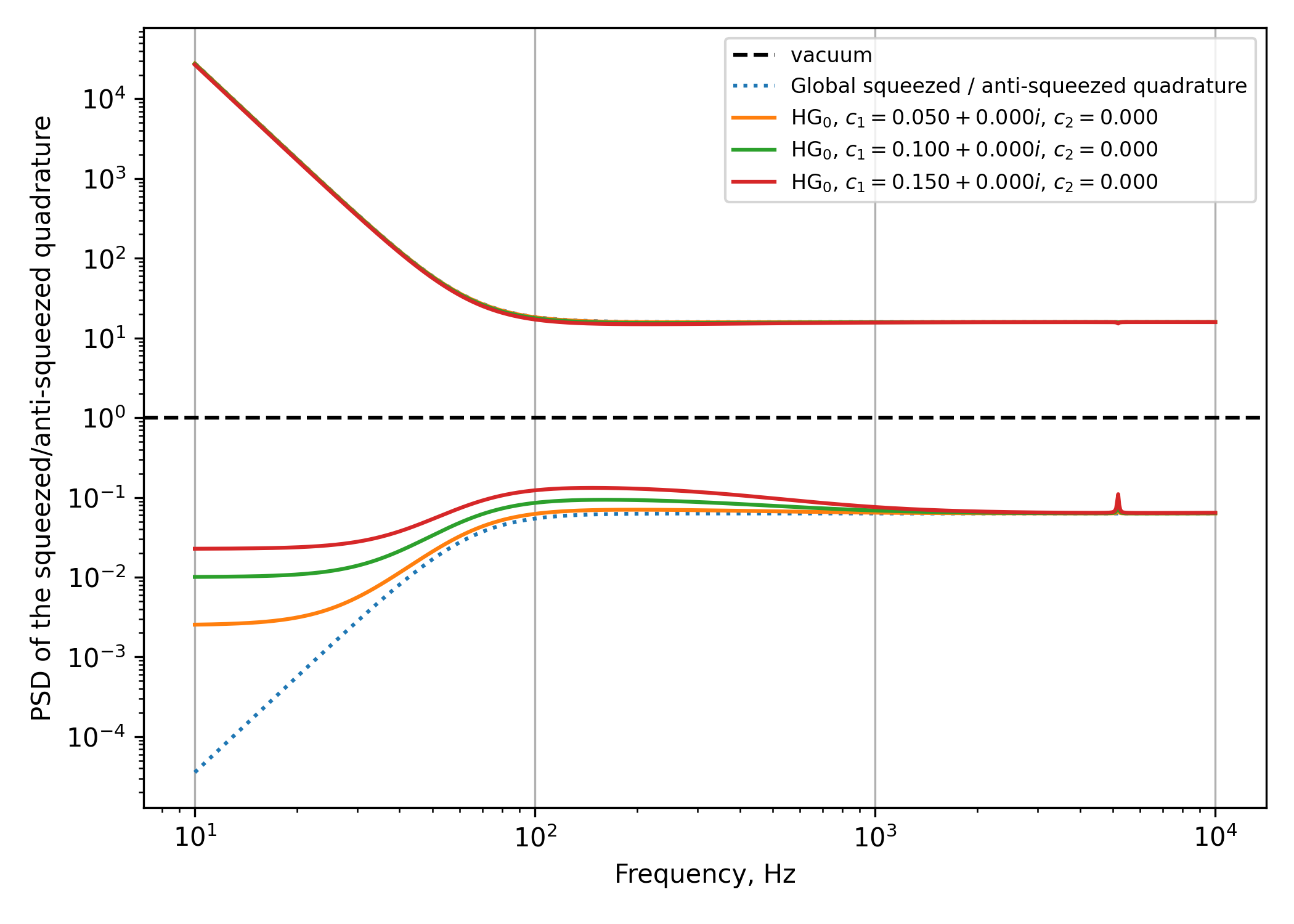}
    \caption{Real $C_1\neq0$, $C_2=0$.}
    \label{sq2}
\end{subfigure}

\vspace{0.3cm}

\begin{subfigure}{0.48\textwidth}
    \centering
    \includegraphics[width=\linewidth]{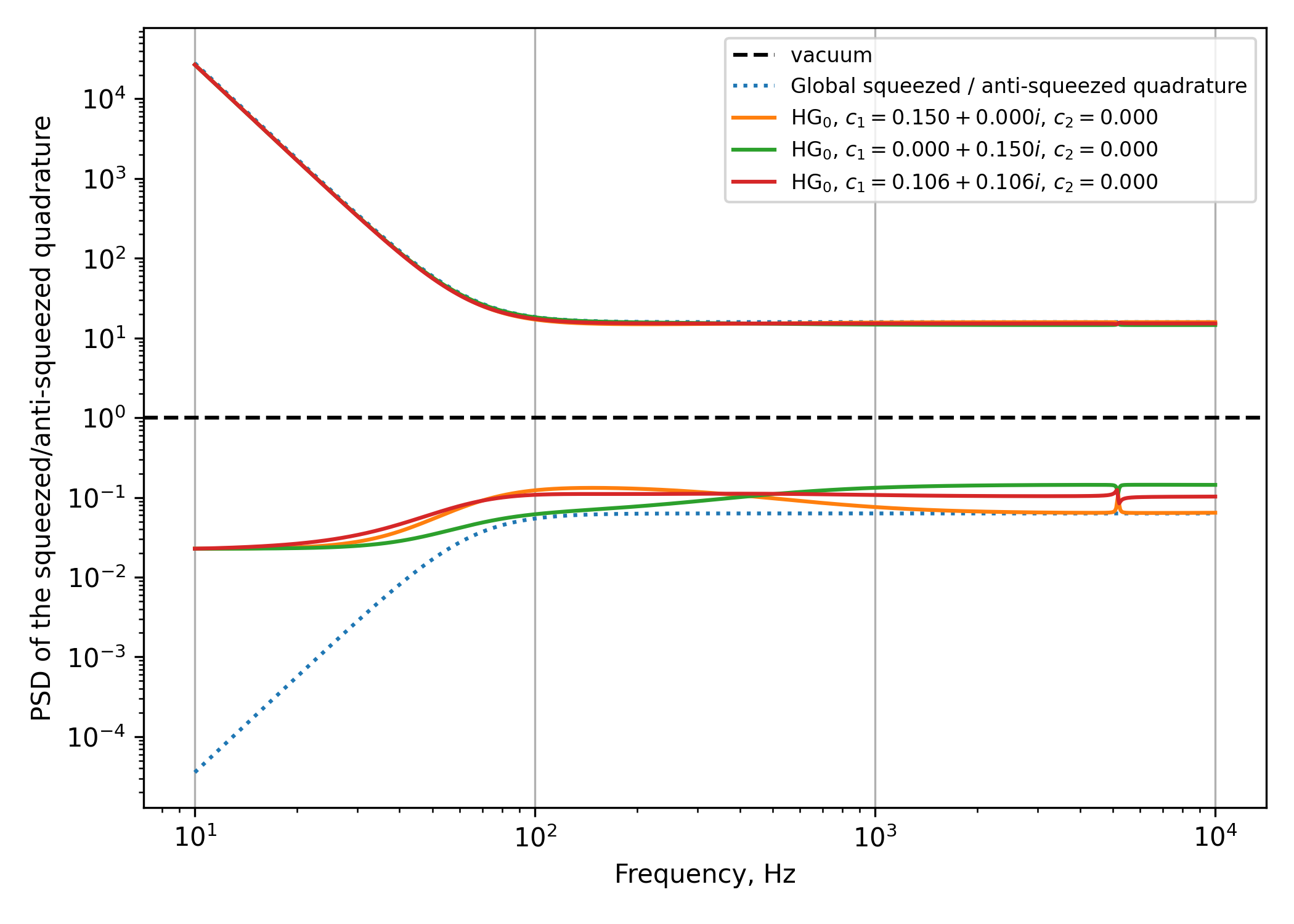}
    \caption{Complex $C_1\neq0$, $C_2=0$.}
    \label{sq3}
\end{subfigure}
\hfill
\begin{subfigure}{0.48\textwidth}
    \centering
    \includegraphics[width=\linewidth]{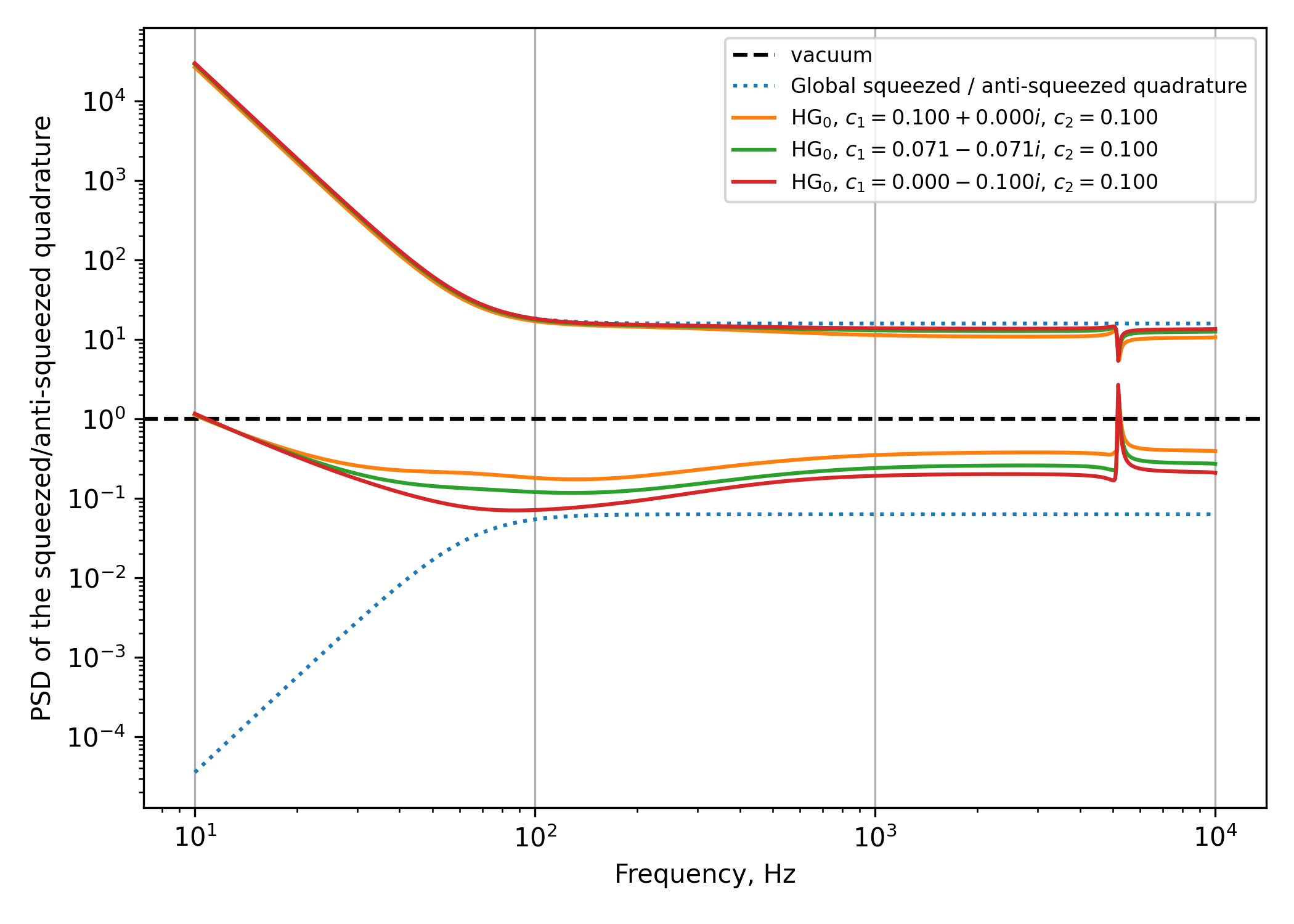}
    \caption{$C_1\neq0$, $C_2\neq0$.}
    \label{sq4}
\end{subfigure}

\caption{
Spectral densities of the squeezed and anti-squeezed quadratures of the
fundamental spatial mode for different mode-mismatch configurations.
The dashed curves correspond to the perfectly matched interferometer.
Mode mismatch reduces the observable squeezing of the fundamental mode,
with the strongest effect produced by the parameter $C_2$.
}
\label{fundamental_squeezing}
\end{figure}

The spectral densities of the squeezed and anti-squeezed quadratures of the
fundamental mode are shown in Fig.~\ref{fundamental_squeezing} for different
values of the mismatch parameters $C_1$ and $C_2$. The dashed curves correspond
to the perfectly matched interferometer and are shown for comparison.

The results show that mode mismatch generally reduces the observable squeezing
of the fundamental mode. As the mismatch increases, the minimum quadrature
spectral density approaches the vacuum level, while the anti-squeezed quadrature
is modified accordingly. This degradation is significantly more pronounced for
variations of $C_2$ than for variations of $C_1$, in agreement with the
sensitivity curves discussed in Sec.~\ref{qn_sensitivity_mismatch}.

At low frequencies, both the squeezed and anti-squeezed quadrature spectral densities deviate significantly from their high-frequency values. This behavior is not directly related to mode mismatch and is already present in the perfectly matched interferometer. It originates from the optomechanical interaction between the optical field and the test masses. Radiation-pressure fluctuations acting on the mirrors generate quantum back-action, which couples the phase and amplitude quadratures of the intracavity field. As a result, the close to amplitude quadrature becomes squeezed at low frequencies, leading to an increase in both the observed squeezing and anti-squeezing levels. This effect is a well-known manifestation of ponderomotive squeezing
produced by radiation-pressure-induced optomechanical coupling
\cite{Corbitt2006} and is present even in the absence of the spatial
mode mixing considered in this work.

\begin{figure}[t]
\centering

\begin{subfigure}{0.48\textwidth}
    \centering
    \includegraphics[width=\linewidth]{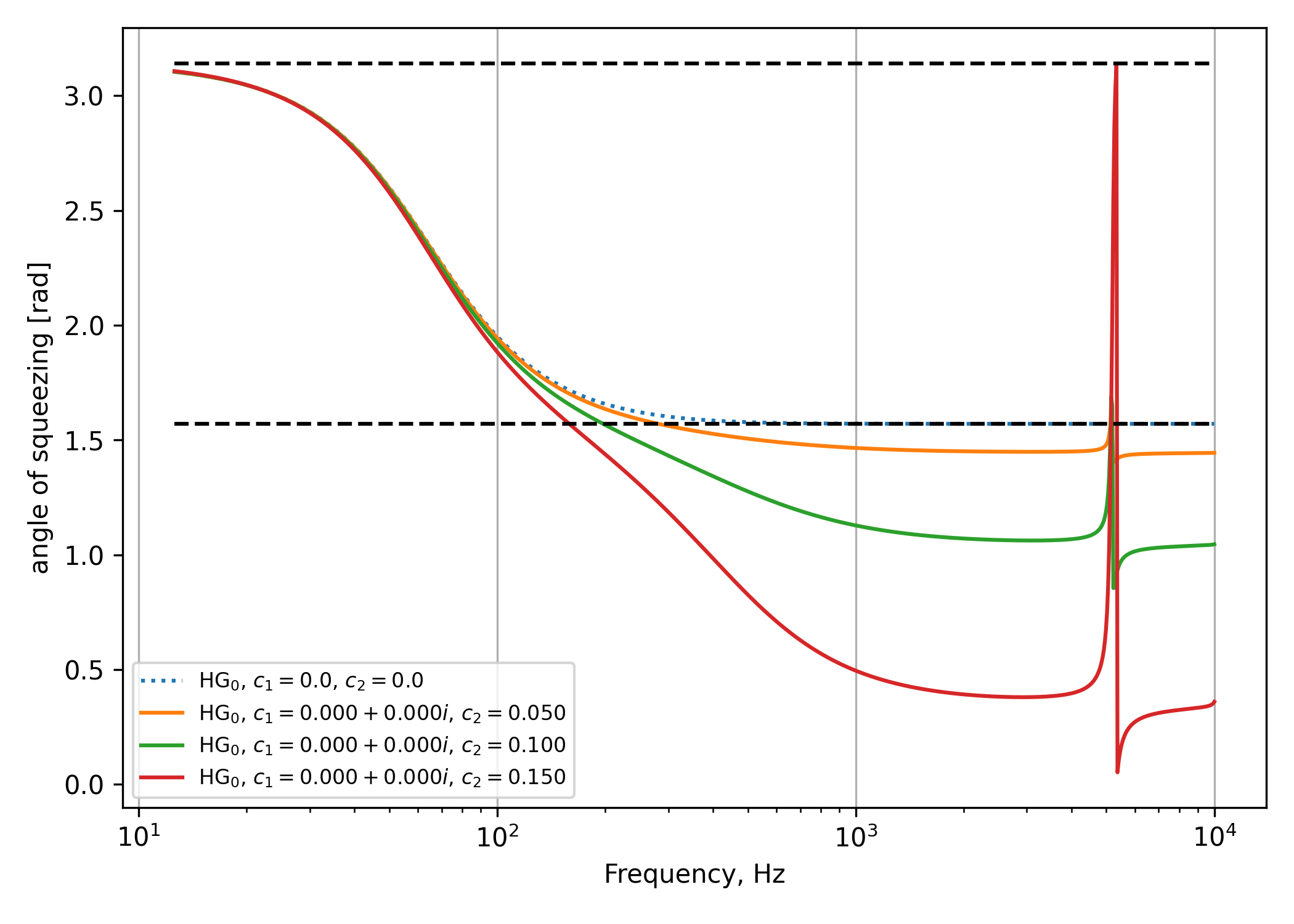}
    \caption{$C_1=0$, $C_2\neq0$.}
    \label{ang1}
\end{subfigure}
\hfill
\begin{subfigure}{0.48\textwidth}
    \centering
    \includegraphics[width=\linewidth]{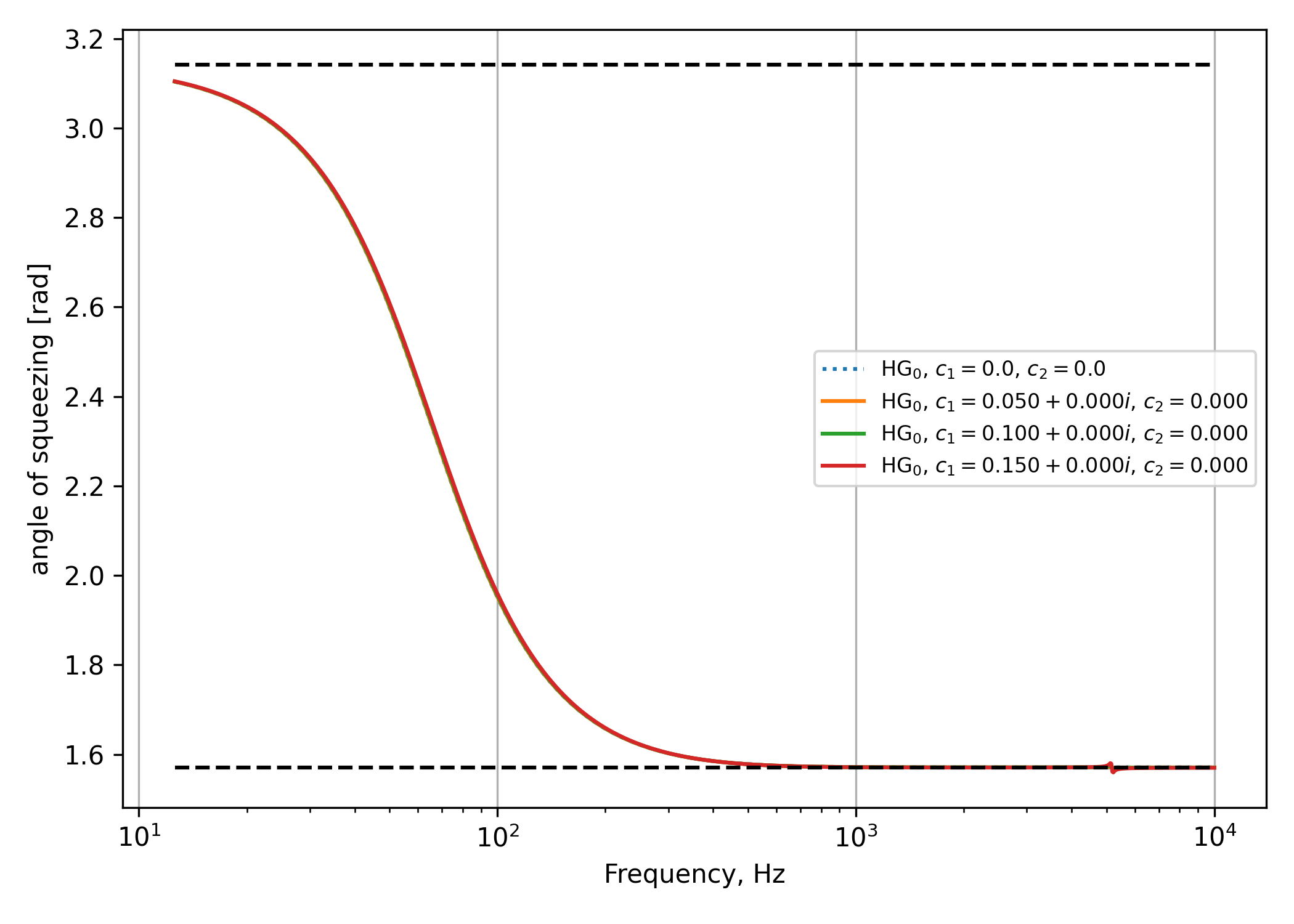}
    \caption{Real $C_1\neq0$, $C_2=0$.}
    \label{ang2}
\end{subfigure}

\vspace{0.3cm}

\begin{subfigure}{0.48\textwidth}
    \centering
    \includegraphics[width=\linewidth]{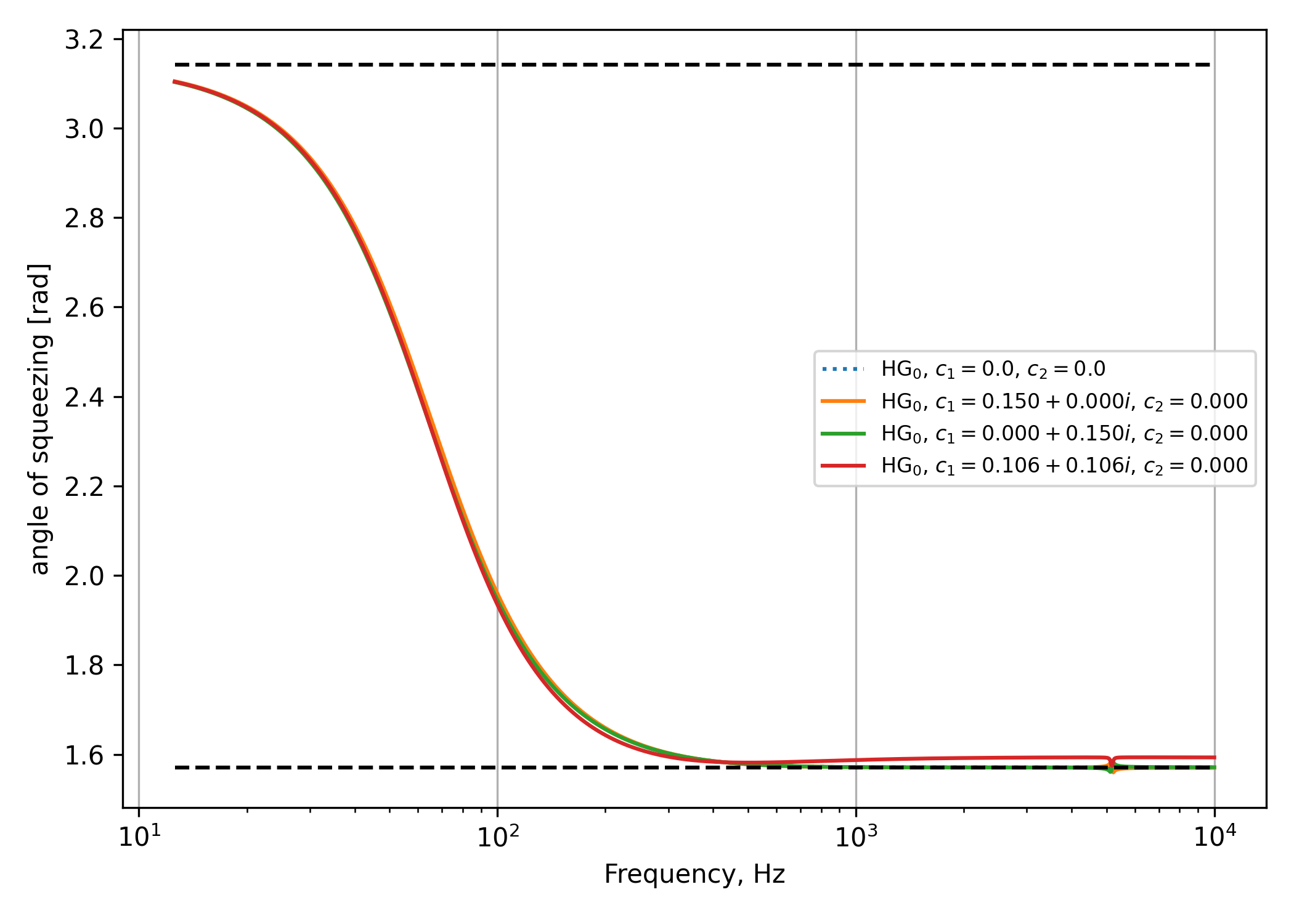}
    \caption{Complex $C_1\neq0$, $C_2=0$.}
    \label{ang3}
\end{subfigure}
\hfill
\begin{subfigure}{0.48\textwidth}
    \centering
    \includegraphics[width=\linewidth]{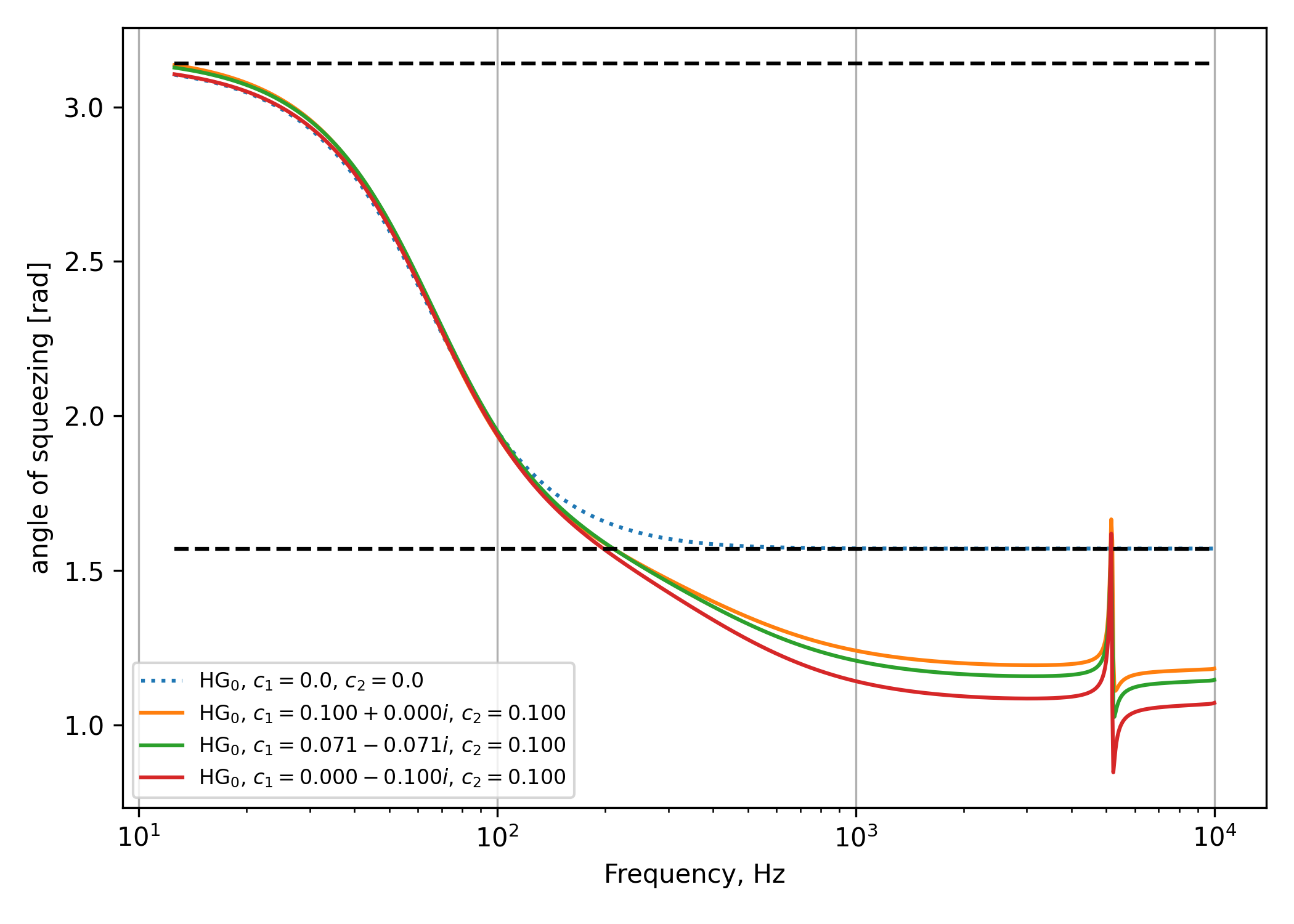}
    \caption{$C_1\neq0$, $C_2\neq0$.}
    \label{ang4}
\end{subfigure}

\caption{
Frequency dependence of the squeezing angle of the fundamental spatial mode
for different mismatch configurations.
The parameter $C_1$ produces only a weak rotation of the squeezing ellipse,
whereas $C_2$ leads to a pronounced frequency-dependent rotation,
particularly in the vicinity of higher-order-mode resonances.
}
\label{fundamental_squeezing_angle}
\end{figure}

Mode mismatch affects not only the magnitude of squeezing but also the frequency-dependent
orientation of the squeezing ellipse.  The frequency dependence of the
fundamental-mode squeezing angle is shown in
Fig.~\ref{fundamental_squeezing_angle}. The parameter $C_1$ produces only a
weak rotation of the squeezing ellipse over the detection band. In contrast,
$C_2$ induces a substantial frequency-dependent rotation, which becomes
especially pronounced near the resonances of higher-order spatial modes.

This behavior provides a natural explanation for the sensitivity modification
observed in the previous section. In the present model, the injected squeezing
is frequency independent, and the detected quadrature is not optimal at all
frequencies. Therefore, the mismatch-induced rotation of the squeezing ellipse
may accidentally move the measured quadrature closer to the optimal one,
leading to a local improvement of sensitivity despite the overall degradation
of the observable squeezing in the fundamental mode.

However, the degradation of the fundamental-mode squeezing does not mean that
the injected squeezing is destroyed. The full optical evolution remains
unitary, so the quantum information carried by the squeezed state cannot simply
disappear. Instead, the squeezing is redistributed among the coupled spatial
modes. This indicates the emergence of nontrivial quantum correlations between
the fundamental mode and higher-order modes, which are analyzed in the next
section.

\subsection{Effective SEC detuning and its compensation}
\label{effective_SEC_detuning}

\begin{figure*}[t]
    \centering

    \begin{subfigure}[t]{0.49\textwidth}
        \centering
        \includegraphics[width=\linewidth]
        {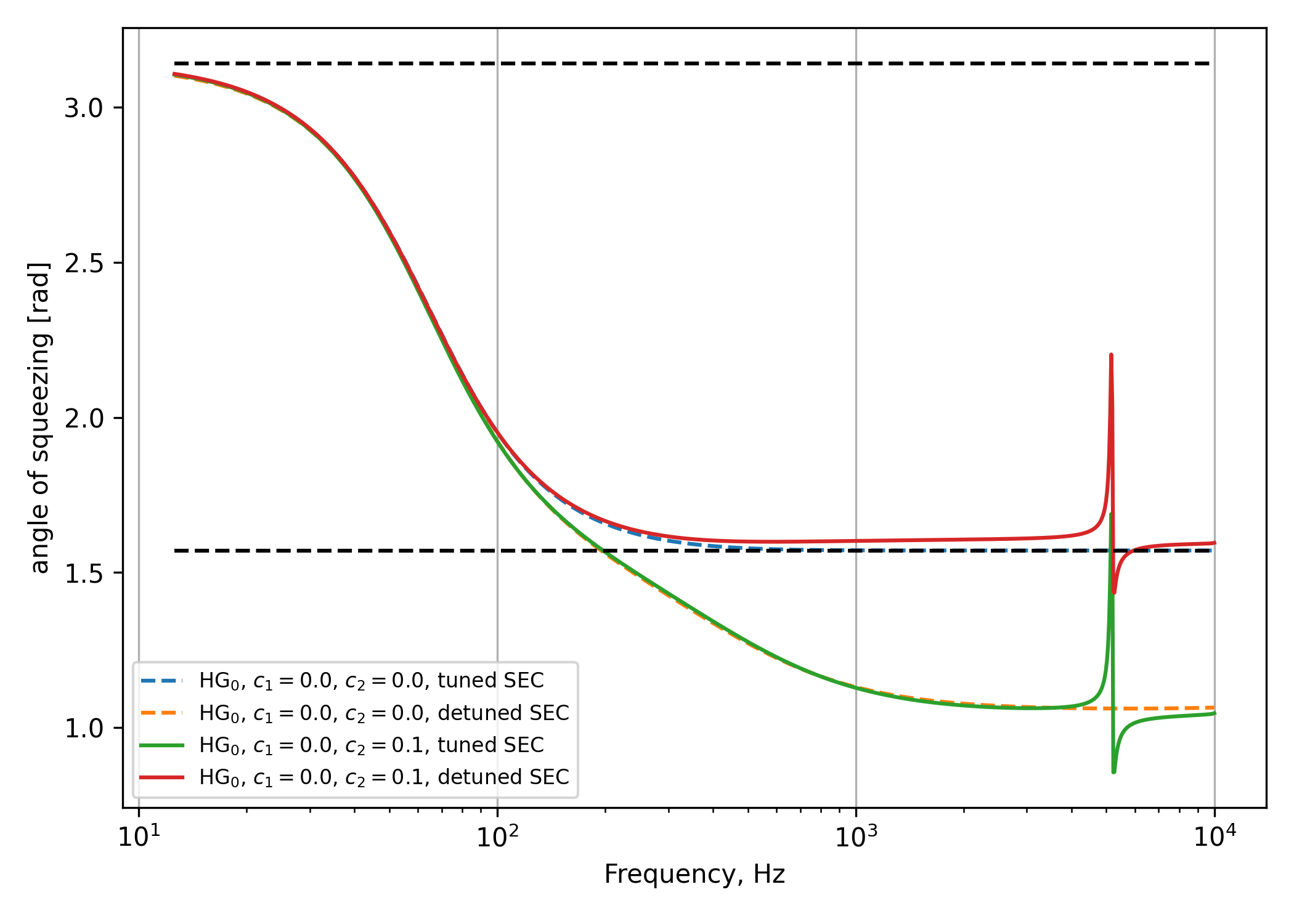}
        \caption{}
        \label{sec_detuning_angle}
    \end{subfigure}
    \hfill
    \begin{subfigure}[t]{0.49\textwidth}
        \centering
        \includegraphics[width=\linewidth]
        {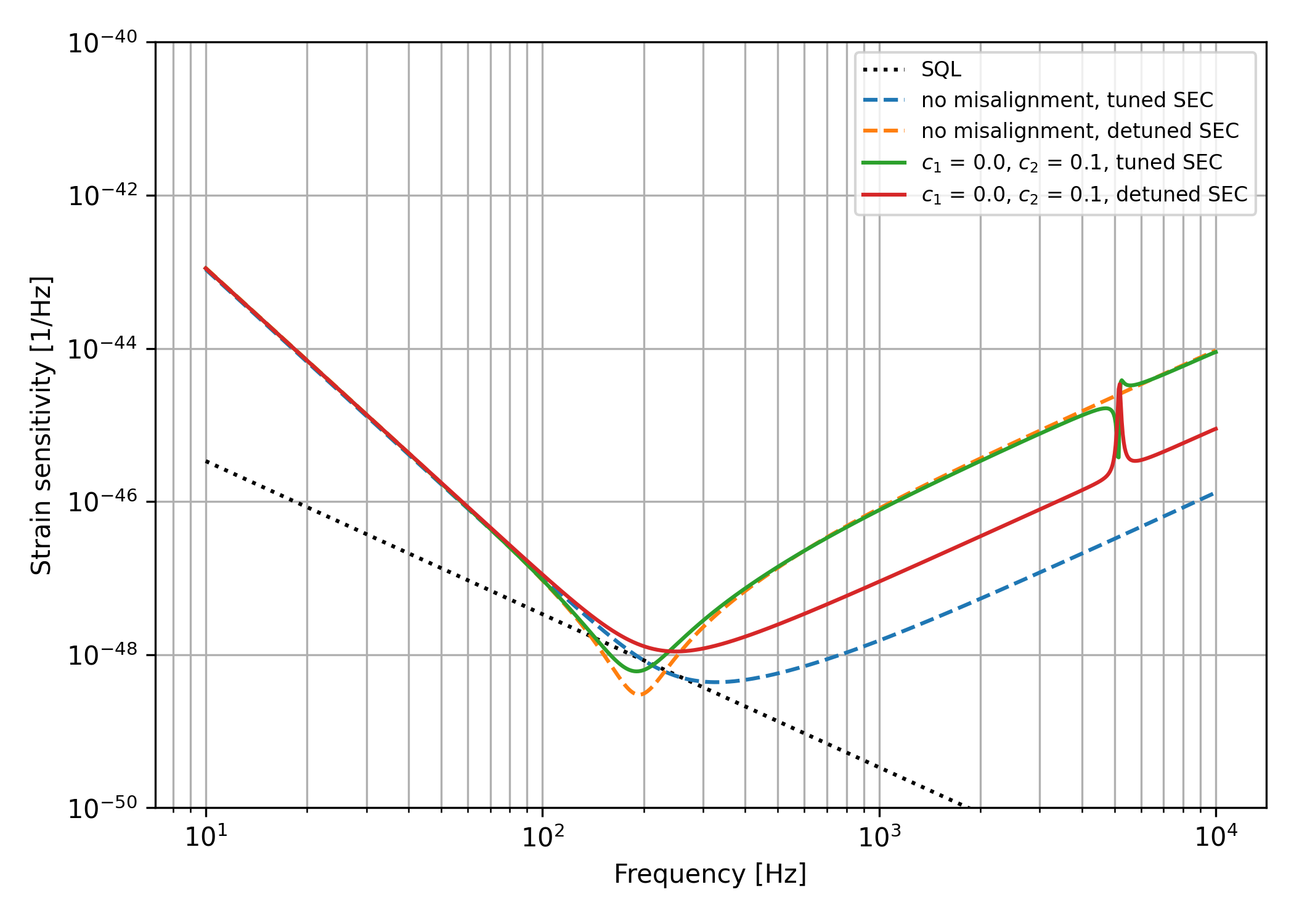}
        \caption{}
        \label{sec_detuning_sensitivity}
    \end{subfigure}

    \caption{
    Effect of compensating the effective SEC detuning induced by internal
    mode mismatch.
    (a) Squeezing angle of the fundamental spatial mode for four
    configurations: an aligned interferometer with zero SEC phase offset,
    an aligned interferometer with an intentional SEC phase offset, an
    interferometer with internal mode mismatch ($C_2=0.1$) and zero SEC
    phase offset, and the same mismatched interferometer after introducing
    a compensating SEC phase offset. The smooth broadband rotation induced
    by mode mismatch is largely removed by the compensating SEC phase
    offset, whereas the narrow resonant feature associated with higher-order
    spatial modes remains.
    (b) Corresponding strain sensitivities. Compensation of the effective
    SEC detuning substantially improves the high-frequency sensitivity but
    does not remove the residual degradation caused by coherent multimode
    coupling.
    }
    \label{sec_detuning_compensation}
\end{figure*}

The results presented in the previous sections were obtained assuming zero
SEC phase offset. Recently, Kuns and Brown~\cite{26KunsArxiv} showed that internal mode
mismatch between the signal extraction cavity and the arm cavities can itself
induce an effective SEC detuning. In their model, coupling of the fundamental
mode to higher-order spatial modes introduces additional Gouy phase shifts
that modify the effective complex reflectivity of the coupled cavity.
As a result, the interferometer behaves as if the SEC were detuned, even
though no intentional SEC phase offset is introduced.

The effect of an intentional SEC phase offset on squeezed-light injection
was previously analyzed by McCuller \textit{et al.}~\cite{21McCullerPRD}.
They showed that SEC detuning produces a characteristic smooth broadband
rotation of the squeezing ellipse. Therefore, if the
mechanism proposed by Kuns and Brown contributes significantly to the
degradation observed in our calculations, the squeezing-angle rotation
produced by internal mode mismatch should resemble that produced by an
intentionally introduced SEC phase offset.

To illustrate this mechanism numerically within our model, additional
simulations were performed for the representative case of $C_2=0.1$.
Figure~\ref{sec_detuning_compensation}(a) compares the squeezing angle
of the fundamental mode for four configurations:
(i) an aligned interferometer with zero SEC phase offset,
(ii) an aligned interferometer with an intentional SEC phase offset,
(iii) an interferometer with internal mode mismatch ($C_2=0.1$) and zero
SEC phase offset, and
(iv) the same mismatched interferometer after introducing a compensating
SEC phase offset.

The smooth broadband rotation produced by internal mode mismatch closely
reproduces the rotation obtained by introducing an intentional SEC phase
offset in the aligned interferometer. Moreover, the compensating SEC phase
offset largely removes this broadband component while leaving the narrow
resonant feature associated with higher-order spatial modes almost unchanged.
These results provide direct numerical support for the interpretation
proposed by Kuns and Brown~\cite{26KunsArxiv}: a substantial fraction of
the squeezing-angle rotation caused by internal mode mismatch originates
from an effective SEC detuning.

The corresponding strain sensitivities are shown in
Fig.~\ref{sec_detuning_compensation}(b). Compensation of the effective
SEC detuning significantly improves the high-frequency sensitivity compared
with the uncompensated mismatched configuration. However, the compensated
sensitivity does not fully recover that of the perfectly aligned
interferometer. In particular, the resonant feature associated with
higher-order spatial modes remains essentially unchanged. This demonstrates
that adjusting the SEC phase offset compensates only the detuning-like
contribution, whereas the remaining degradation originates from coherent
multimode coupling.

The purpose of this analysis is to clarify the physical origin of the
degradation caused by internal mode mismatch. The results presented in
Figs.~\ref{strain_sensitivity_mismatch} - \ref{fundamental_squeezing_angle} therefore represent the complete response of the interferometer
to internal mode mismatch when the SEC phase offset is zero. The present
analysis shows that this response consists of two distinct physical
contributions: a smooth broadband component originating from the induced
effective SEC detuning, which can be substantially mitigated by an
appropriate SEC phase offset, and a residual resonant component arising from
coherent multimode interaction that cannot be removed by a single adjustment
of the SEC phase.

\subsection{Multimode quantum correlations and spatial-mode entanglement}
\label{multimode_entanglement}

The results presented in the previous subsection \ref{fundamental_mode_squeezing}
demonstrate that mode mismatch
leads to a substantial degradation of the observable squeezing in the
fundamental spatial mode. At first sight, this behavior may suggest that the
injected squeezed state is partially destroyed by mode mismatch. However, such
an interpretation would contradict the unitary nature of the optical evolution
considered in this work. Since the interferometer is assumed to be lossless,
the quantum information carried by the injected squeezed state cannot
disappear. Instead, it must be redistributed among the coupled spatial modes.

To investigate this redistribution, we analyze the full multimode spectral
density matrix $S_1$ \eqref{eq:output_spectral_matrix_main}. While the observable squeezing of the fundamental mode
decreases with increasing mismatch, the minimum eigenvalue of the complete
multimode covariance matrix remains close to the value obtained in the
perfectly matched interferometer. This value is shown by the blue dashed
curves in Fig.~\ref{fundamental_squeezing}, which correspond
simultaneously to the squeezed quadrature of the fundamental mode in the
absence of mismatch and to the minimum noise achievable by an optimal
multimode quadrature in the mismatched interferometer.

This result demonstrates that the injected squeezing is preserved within the
optical field but is no longer associated with a single spatial mode.
Instead, the most strongly squeezed quadrature becomes a
frequency-dependent linear combination of quadratures belonging to several
spatial modes,

\begin{equation}
X_{\mathrm{opt}}(\Omega)
=
\sum_n
\left[
u_n(\Omega) a_n^a
+
v_n(\Omega) a^{\phi}_n
\right].
\end{equation}

Consequently, mode mismatch transforms the initially single-mode squeezed
state into a genuinely multimode squeezed state. The apparent loss of
squeezing observed in the fundamental mode therefore reflects a
redistribution of squeezing among spatial modes rather than a destruction of
the squeezed state itself.

The redistribution of squeezing is accompanied by the appearance of
correlations between spatial modes. To quantify these correlations, we
analyze the off-diagonal blocks of the multimode spectral density matrix.
In the absence of mismatch, the fundamental mode and higher-order modes are
independent and the corresponding cross-correlations vanish. As the mismatch
parameters increase, nonzero correlations emerge between the fundamental
mode and higher-order modes. Their magnitude grows with increasing mismatch,
indicating progressively stronger coupling between the spatial modes.

The appearance of intermode correlations demonstrates that the optical field
can no longer be described as a product of independent single-mode states.
However, the existence of correlations alone does not prove the presence of
quantum entanglement, since classical correlations may produce a similar
signature. To distinguish between classical and quantum correlations, we
apply the positive partial transpose (PPT) criterion for Gaussian states
\cite{Simon2000}.

The fundamental mode is treated as one subsystem, while all higher-order
modes form the second subsystem. For each frequency, we evaluate the minimum eigenvalue of

\begin{equation}
S_1^\Gamma(\Omega)
+
i\Omega_s,
\end{equation}

where $S_1^\Gamma(\Omega)$ is the partially transposed spectral density
matrix and $\Omega_s$ is the symplectic form of the multimode quadrature
space.

For Gaussian states, the partial transposition operation has a simple
representation in phase space: it changes the sign of the momentum quadratures
of the transposed subsystem. Therefore, for the spectral density matrix
$S_1(\Omega)$ \eqref{eq:output_spectral_matrix_main}, the partially transposed matrix is written as
\begin{equation}
S_1^\Gamma(\Omega)
=
\Lambda S_1(\Omega)\Lambda ,
\end{equation}
where, for the bipartition HG$_0$ : HOM, the transformation matrix has the form
\begin{equation}
\Lambda
=
\mathrm{diag}
\left(
1,\,
1,\,
1,\,
-1,\,
1,\,
-1,\,
\dots,\,
1,\,
-1
\right).
\end{equation}
Here the first two entries correspond to the fundamental mode and are left
unchanged, whereas for each higher-order mode the phase quadrature changes
sign.

The PPT criterion states that any separable Gaussian state must satisfy

\begin{equation}
S_1^\Gamma(\Omega)
+
i\Omega_s
\ge 0.
\end{equation}

Therefore, a negative minimum eigenvalue

\begin{equation}
\lambda_{\min}
\left[
S_1^\Gamma(\Omega)
+
i\Omega_s
\right]
<
0
\end{equation}
provides a sufficient signature of nonseparability and therefore of quantum
entanglement between the fundamental mode and the subsystem of higher-order
modes.

\begin{figure}[t]
\centering

\begin{subfigure}{0.48\textwidth}
    \centering
    \includegraphics[width=\linewidth]{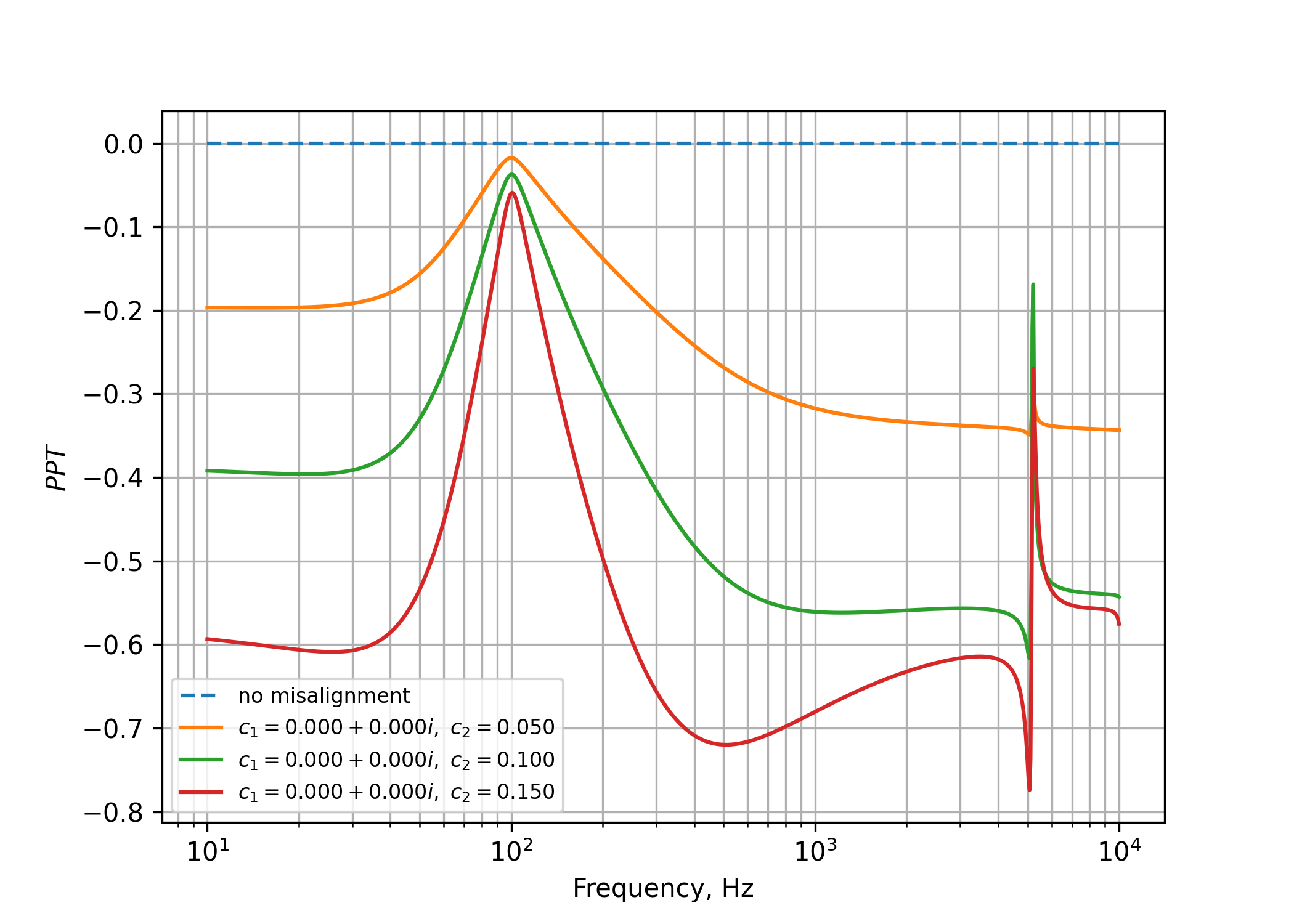}
    \caption{$C_1=0$, $C_2\neq0$.}
    \label{ppt1}
\end{subfigure}
\hfill
\begin{subfigure}{0.48\textwidth}
    \centering
    \includegraphics[width=\linewidth]{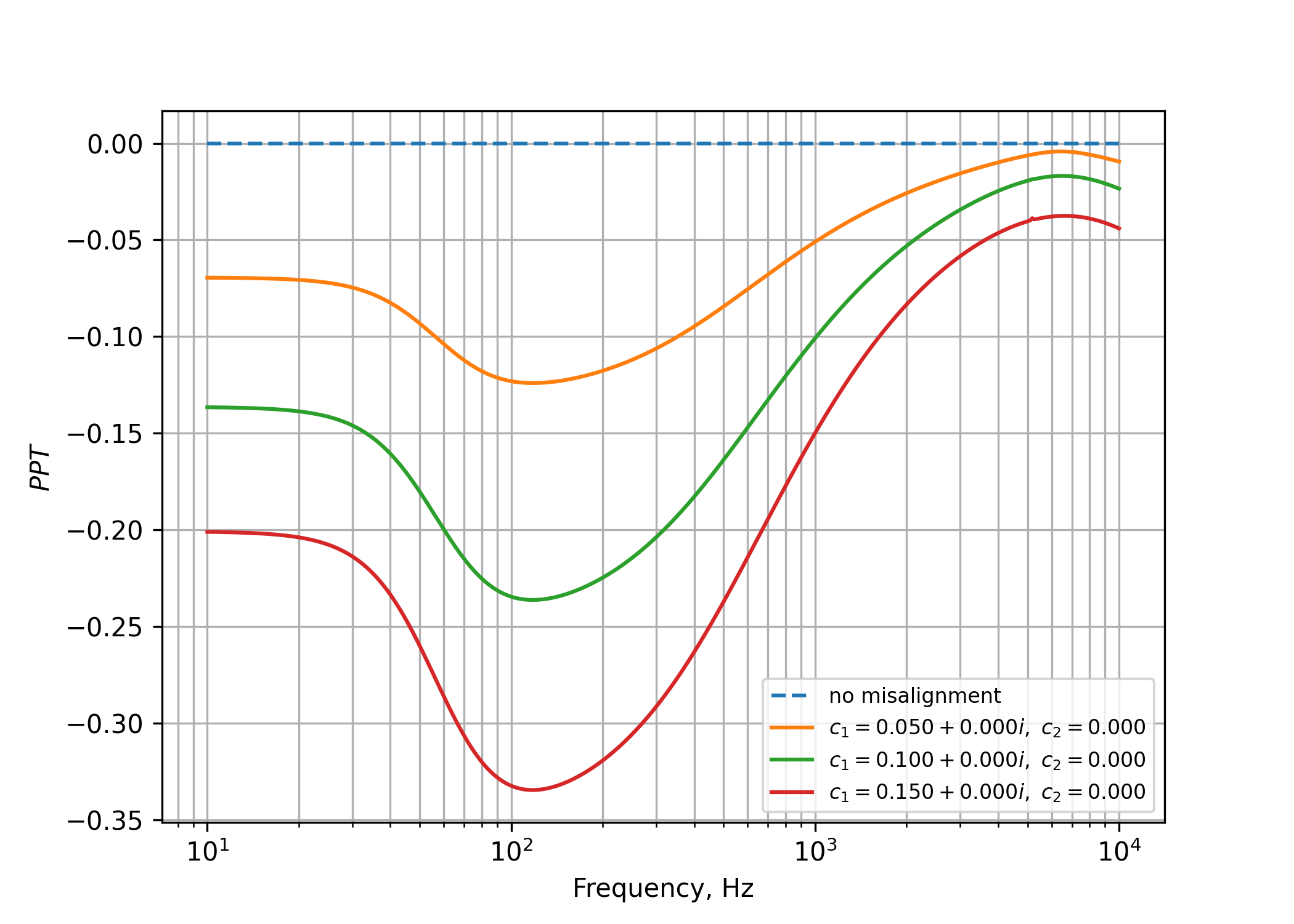}
    \caption{Real $C_1\neq0$, $C_2=0$.}
    \label{ppt2}
\end{subfigure}

\vspace{0.3cm}

\begin{subfigure}{0.48\textwidth}
    \centering
    \includegraphics[width=\linewidth]{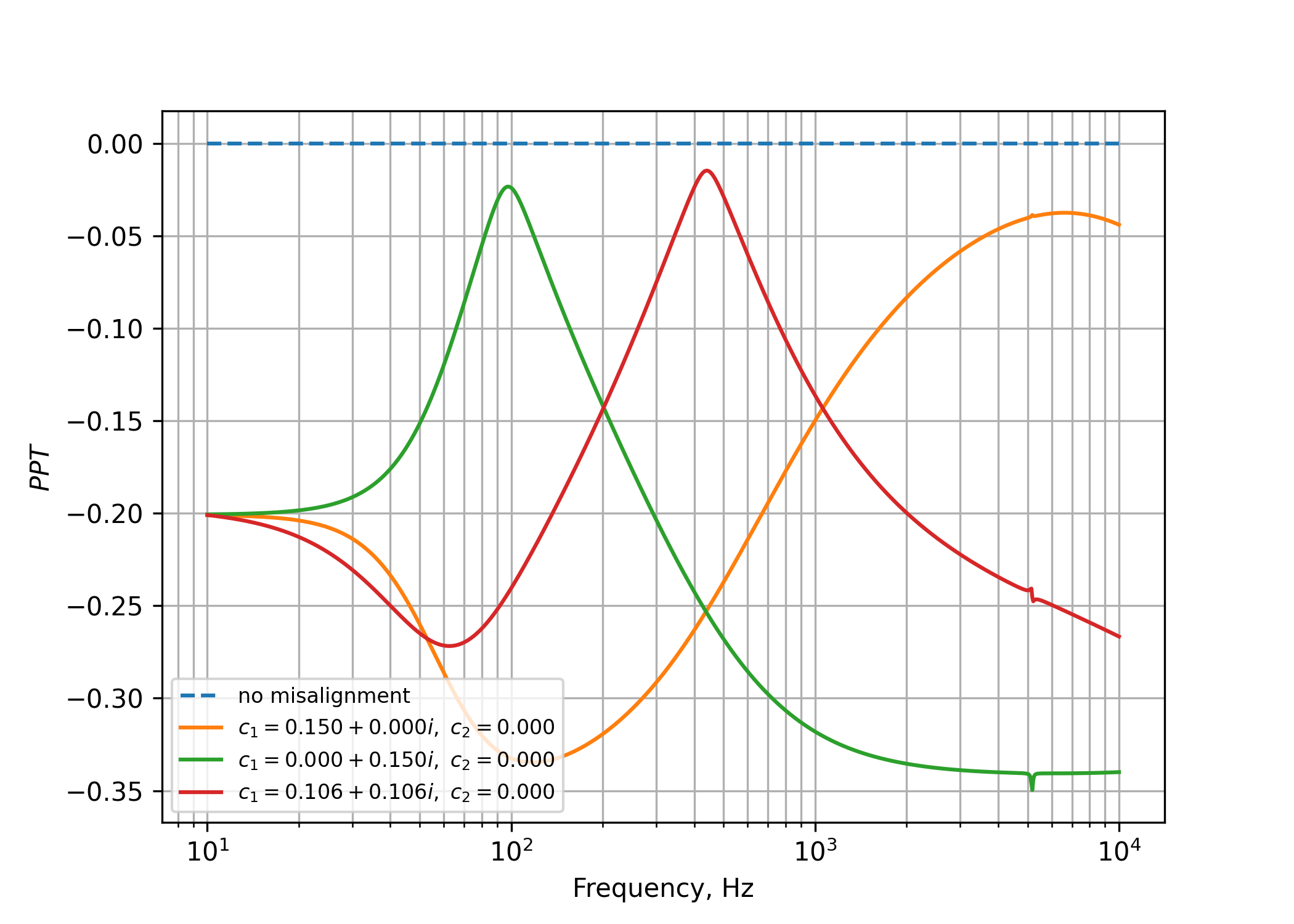}
    \caption{Complex $C_1\neq0$, $C_2=0$.}
    \label{ppt3}
\end{subfigure}
\hfill
\begin{subfigure}{0.48\textwidth}
    \centering
    \includegraphics[width=\linewidth]{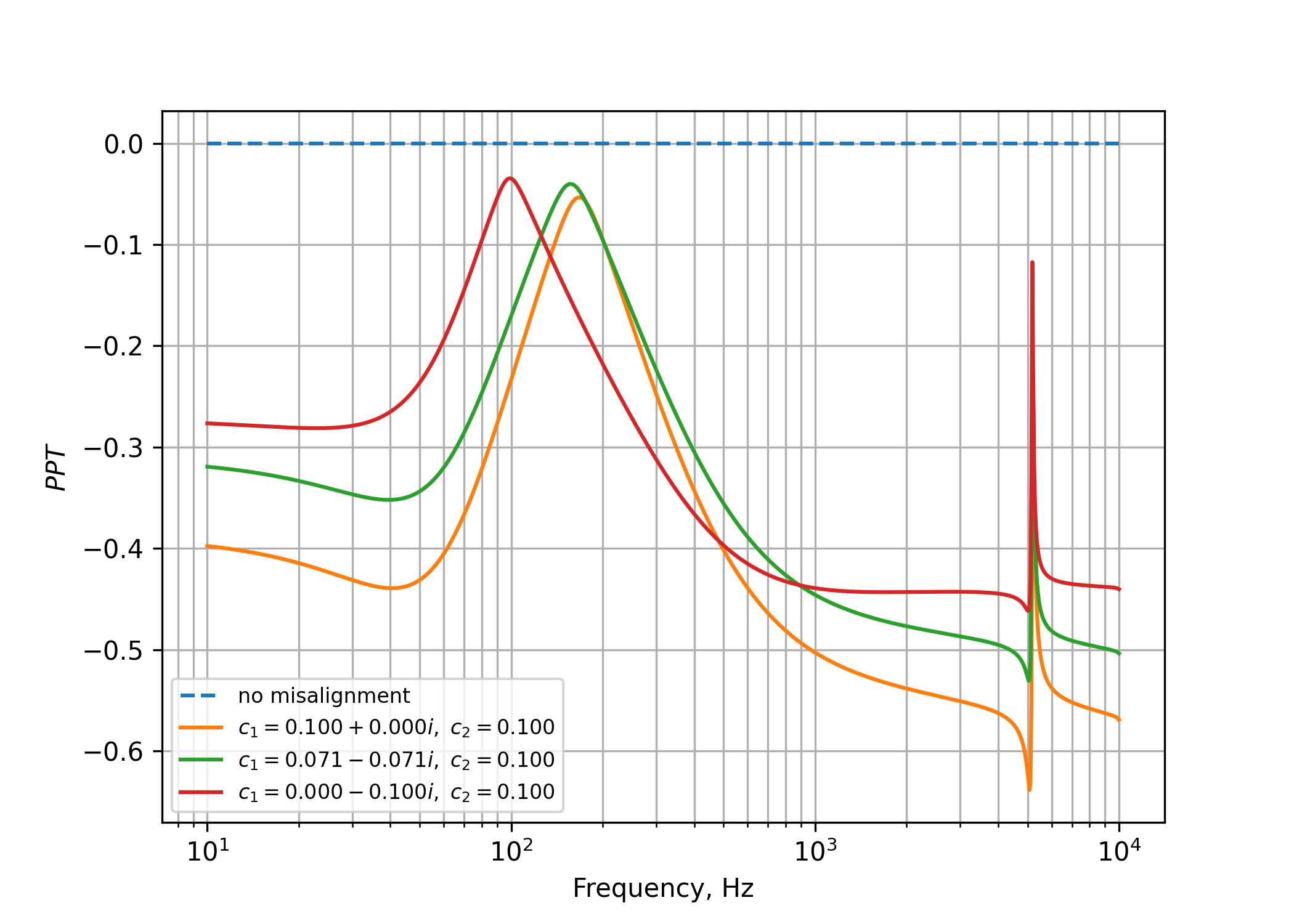}
    \caption{$C_1\neq0$, $C_2\neq0$.}
    \label{ppt4}
\end{subfigure}

\caption{
Minimum eigenvalue
$\lambda_{\min}\!\left[S_1^\Gamma(\Omega)+i\Omega_s\right]$
for different mode-mismatch configurations.
Negative values indicate violation of the PPT criterion and therefore
nonseparability of the bipartition HG$_0$ : HOM.
The magnitude of the violation increases with increasing mode mismatch,
demonstrating the growth of quantum correlations between the fundamental
mode and higher-order spatial modes. The PPT criterion is evaluated for the bipartition consisting of the
fundamental spatial mode HG$_0$ and the subsystem formed by all
higher-order modes.
}
\label{ppt}
\end{figure}

The results of the PPT analysis are shown in
Fig.~\ref{ppt}. Significant violations of the PPT criterion are observed
over a broad frequency range whenever mode mismatch is present.
Furthermore, the magnitude of the violation increases with increasing
mismatch, demonstrating that the strength of the generated quantum
correlations grows together with the mode coupling.

These results establish a direct connection between mode mismatch,
multimode squeezing, and spatial-mode entanglement. Rather than destroying
the injected squeezing, mode mismatch redistributes it among several
spatial modes and generates quantum entanglement between them. The apparent
degradation of squeezing observed in the fundamental mode is therefore a
manifestation of the transition from a single-mode squeezed state to a
multimode entangled optical state.

The physical picture emerging from our analysis can be summarized as
follows. Optical-axis mismatch introduces coherent coupling between
different spatial modes. This coupling generates intermode correlations and
redistributes the injected squeezing among several modes. As a result, the
squeezed state becomes intrinsically multimode, and the fundamental mode
becomes entangled with the subsystem of higher-order modes. The degradation
of the observable squeezing in the fundamental mode is therefore not caused
by the loss of quantum information but rather by its redistribution within
the multi-mode optical field.

It should be emphasized that the interferometer model considered in this work
describes a closed multimode optical system in which optical-axis misalignment
results only in coherent coupling between spatial modes. All optical elements
are assumed to have infinite transverse apertures, and therefore diffraction
and clipping losses are neglected. No other optical losses are included in the
present model. Within this framework, the degradation of the observable squeezing
in the fundamental mode should not be interpreted as an irreversible loss of the injected quantum resource. Instead,
the squeezing is redistributed among the interacting spatial modes.

This observation suggests a possible route for mitigating the adverse effects of misalignment. If the slowly varying mismatch parameters $C_1$ and $C_2$ can be monitored in real time, it may become possible to adapt the readout scheme to the instantaneous multimode structure of the optical field and measure a near-optimal combination of quadratures involving both the fundamental and higher-order modes. Such an adaptive strategy could be combined with compensation of the effective SEC detuning discussed in Sec.~\ref{effective_SEC_detuning}. At the same time, improving the alignment control system to directly reduce the mismatch parameters is expected to remain the primary mitigation strategy. Nevertheless, practical alignment servos cannot completely eliminate residual time-dependent mode mismatch, suggesting that adaptive readout or SEC tuning may provide an additional improvement under realistic operating conditions.

However, the present analysis is restricted to the quantum noise field and does not include the influence of misalignment on the gravitational-wave signal itself. Moreover, realistic interferometers are subject to optical losses, which may substantially modify the conclusions obtained within the idealized lossless model considered here. Investigating these effects will be the subject of future work.

\section{Conclusion}

In this work, we developed a multimode matrix formalism describing
optical-axis misalignment in the differential mode of a signal-extracted gravitational-wave interferometer. The differential optical mode was represented as a system of two coupled resonators corresponding to the signal extraction cavity and the arm cavity. Within this framework, analytical expressions were derived that directly relate geometrical perturbations of optical elements to the coefficients of the mode-mixing matrix. This establishes a direct connection between mechanical
imperfections of the interferometer and the resulting multimode optical dynamics.

Using the developed model, we investigated the influence of optical-axis misalignment on the quantum-noise-limited sensitivity of the interferometer. The obtained results demonstrate that different types of misalignment affect the detector performance in qualitatively different
ways. In particular, mismatch between the signal extraction cavity and the arm cavity produces significantly stronger degradation of sensitivity than mismatch between the readout system and the signal extraction cavity. The resulting mode coupling also gives rise to resonant spectral features associated with higher-order spatial modes.

To understand the origin of this degradation, we analyzed the squeezing
properties of the optical field. We showed that optical-axis misalignment
leads both to a reduction of the observable squeezing in the fundamental
spatial mode and to a frequency-dependent rotation of the squeezing
ellipse. A substantial part of the broadband squeezing-angle rotation produced by
the internal SEC--arm-cavity mismatch was found to originate from an
effective SEC detuning~\cite{26KunsArxiv}. Introducing an appropriate SEC phase offset largely
compensates this broadband rotation and significantly improves the
high-frequency sensitivity. However, the resonant features associated with
higher-order spatial modes remain after compensation, demonstrating that
the mismatch-induced response contains both a detuning-like contribution
and a residual contribution arising from coherent multimode coupling.

However, within the idealized lossless model considered here, the observed
degradation of the fundamental-mode squeezing does not imply the destruction
of the injected quantum resource. Since finite apertures and other optical
losses are neglected, the optical evolution remains unitary. The squeezed state is therefore preserved within
the optical field and redistributed among the coupled spatial modes. The
minimum noise achievable by an optimal multimode quadrature remains close to
that of the perfectly aligned system, indicating that, within this model, the
apparent loss of squeezing originates from a redistribution of quantum
correlations rather than from irreversible optical loss.

The most important consequence of this redistribution is the emergence of quantum correlations between spatial modes. By analyzing the covariance matrix of the multimode optical field, we demonstrated the appearance of strong intermode correlations between the fundamental mode and higher-order modes. Application of the positive partial transpose criterion revealed significant violations of the separability condition over a broad frequency range, providing evidence of spatial-mode entanglement generated by optical-axis misalignment. These results show that misalignment transforms an initially single-mode squeezed state into a multimode entangled quantum state.

Several limitations of the present model should be emphasized. The analysis
was restricted to optical-axis perturbations occurring within a single
transverse plane and assumed spherical optical surfaces. Finite apertures,
diffraction and clipping losses of higher-order modes, and all other optical
losses were neglected. Furthermore, only the transformation of the quantum
noise field was considered, while the influence of misalignment on the
gravitational-wave signal itself was not investigated.

Future work will focus on extending the developed formalism to a
complete three-dimensional description of optical-axis perturbations and on analyzing the influence of misalignment on both the quantum noise and the signal field. Since the injected squeezing is redistributed among spatial modes rather than irreversibly lost, an intriguing possibility is the development of adaptive multimode readout schemes capable of recovering a fraction of the lost quantum-noise suppression. Assessing the feasibility of such approaches remains an important direction for future research.

Overall, the results demonstrate that, within the idealized lossless model
considered here, optical-axis misalignment should not be interpreted merely
as an effective loss of observable squeezing. Instead, coherent mode coupling
can generate multimode quantum correlations and spatial-mode entanglement in gravitational-wave interferometers. Understanding these effects requires a fully multimode quantum description and may become increasingly important for future generations of precision interferometric detectors.

Finally, we emphasize that the present model considers only a restricted class of optical-axis misalignment. The analysis is performed for a planar geometry, in which all optical-axis displacements occur within the $OXZ$ plane, while the optical axis of the injected field is assumed to coincide with the optical axis of the readout system. Under these assumptions, the mode mismatch is characterized by two
parameters, $C_1$ and $C_2$, corresponding to the two interfaces between
the coupled optical subsystems. The parameter $C_1$ is generally complex,
whereas $C_2$ is real.

A complete three-dimensional description of optical-axis misalignment would require additional independent mismatch parameters. In the most general case, six complex coefficients are expected to describe the mode-basis transformations between the injected beam, the signal extraction cavity, the arm cavity, and the readout system, accounting for misalignment independently in the two transverse planes. Extending the present formalism to this fully three-dimensional configuration will be the subject of future work.

\acknowledgments
We are deeply grateful to Stefan Danilishin from Maastricht University for statement of the problem and providing valuable advice during the research. 
We regret that he could not be included in the list of authors of the article.

The research of AVK has been supported by Theoretical Physics and Mathematics Advancement Foundation “BASIS” (Contract No. 25-2-2-28-1). AVK and SPV would like to express their gratitude for the support provided by  the Interdisciplinary Scientific and Educational School of Moscow University ``Fundamental and Applied Space Research'' and by the TAPIR GIFT MSU Support of the California Institute of Technology.  This document has LIGO number P2600387.

\appendix

\section{Derivation of the multimode input-output relations}
\label{app:multimode_IO}

In this Appendix, we present the detailed derivation of the multimode
input-output relations used in Sec.~\ref{sec:matrix_description}.

\subsection{Optical propagation in the differential mode}
\label{app:optical_propagation}

Let $\mathbf A_p$ and $\mathbf A_s$ denote the vectors of input spatial-mode amplitudes
entering the power-recycling cavity and the signal extraction cavity,
respectively. The corresponding intracavity fields satisfy

\begin{align}
\mathbf B_p
&=
t_p\mathbf A_p-r_p\mathbf B_{1p},
\label{eq:app_PR_boundary}
\\
\mathbf B_s
&=
t_s\Theta_1\mathbf A_s-r_s\mathbf B_{1s}.
\label{eq:app_SEC_boundary}
\end{align}

Here $t_p$, $r_p$, $t_s$, and $r_s$ are the amplitude transmissivities
and reflectivities of the power- and signal-recycling mirrors. The matrix
$\Theta_1$ transforms the input spatial basis into the SEC eigenmode
basis.

Propagation from the recycling mirrors to the beamsplitter is described
by the diagonal matrices

\begin{equation}
M_{L_{p,s}}
=
\left\{
\delta_{nm}
\exp\left(-ikL_{p,s}\right)
\exp\left[i(n+1)\Delta\psi_{L_{p,s}}\right]
\right\}.
\label{eq:app_recycling_propagation}
\end{equation}
Here, $L_{p,s}$ denotes the distances from the PRM and SEM to the beamsplitter, respectively,
$k$ is the optical wavenumber, and $\Delta\psi_{L_{p,s}}$ are the Gouy phases
accumulated during propagation. 

The fields immediately before the beamsplitter are therefore

\begin{equation}
\mathbf E_{p,s}
=
M_{L_{p,s}}\mathbf B_{p,s}.
\label{eq:app_fields_before_BS}
\end{equation}

After the beamsplitter, the fields entering the north and east arm
cavities are

\begin{align}
\mathbf F_n
&=
M_{l_n}
\frac{
\mathbf E_p+\mathbf E_s
}{
\sqrt{2}
},
\label{eq:app_north_arm_field}
\\
\mathbf F_e
&=
M_{l_e}
\frac{
\mathbf E_p-\mathbf E_s
}{
\sqrt{2}
},
\label{eq:app_east_arm_field}
\end{align}
where
\begin{equation}
M_{l_{n,e}}
=
\left\{
\delta_{nm}
\exp\left(-ikl_{n,e}\right)
\exp\left[i(n+1)\Delta\psi_{l_{n,e}}\right]
\right\}
\label{eq:app_BS_arm_propagation}
\end{equation}
are the propagation matrices between the beamsplitter and the corresponding
input test masses. Here, $l_{n,e}$ denotes the distances from the beamsplitter to the corresponding input test masses, and $\Delta\psi_{l_{n,e}}$ are the Gouy phases
accumulated during propagation.

The common and differential optical modes are decoupled when the
propagation phases in the two arms are identical. We therefore assume a
symmetric interferometer,
\begin{equation}
l_n=l_e\equiv l,
\qquad
\Delta\psi_{l_n}
=
\Delta\psi_{l_e}
\equiv
\Delta\psi_l,
\label{eq:app_symmetric_arms}
\end{equation}
so that
\begin{equation}
M_{l_n}
=
M_{l_e}
\equiv
M_l.
\label{eq:app_equal_arm_propagation}
\end{equation}
Under this assumption, the common and differential optical modes evolve
independently.

The reflection matrix of the arm cavities is diagonal in the arm-cavity
eigenmode basis and is given by

\begin{equation}
R
=
\left\{
\delta_{nm}
\frac{
\exp\left[2i(n+1)\Delta\psi_L\right]
\exp(-2ikL)
-r_{\mathrm{ITM}}
}{
1
-
r_{\mathrm{ITM}}
\exp\left[2i(n+1)\Delta\psi_L\right]
\exp(-2ikL)
}
\right\}.
\label{eq:app_arm_reflection}
\end{equation}
Here $L$ is the arm length, $\Delta\psi_L$ is the Gouy phases
accumulated during propagation through the arm cavity

The complete round-trip operator of the differential optical mode is

\begin{equation}
S
=
M_{L_s}
M_l
U_2
R
\Theta_2
M_l
M_{L_s}.
\label{eq:app_round_trip}
\end{equation}

The field returning to the SEC in the absence of mirror displacement is
therefore

\begin{equation}
\mathbf B_{1s}
=
S\mathbf B_s.
\label{eq:app_returning_optical_field}
\end{equation}

\subsection{Coupling to the differential mirror displacement}
\label{app:signal_coupling}

The differential displacement of the arm-cavity mirrors is defined as

\begin{equation}
\Delta x(\Omega)
=
x_n(\Omega)-x_e(\Omega).
\label{eq:app_delta_x}
\end{equation}

A differential mirror displacement generates sidebands in the
fundamental arm-cavity mode. Taking this contribution into account, the
field returning to the SEC becomes

\begin{equation}
\mathbf B_{1s}
=
S\mathbf B_s
-
D(\Omega)\Delta x(\Omega),
\label{eq:app_returning_field_with_signal}
\end{equation}
where
\begin{equation}
\mathbf D(\Omega)
=
\sqrt{2}\,i
\frac{
k_0t_{\mathrm{ITM}}C_0
\exp(-i\Omega L/c)
}{
1-r_{\mathrm{ITM}}\exp(-2i\Omega L/c)
}
M_{L_s}M_lU_2\mathbf Y_0.
\label{eq:app_D_vector}
\end{equation}
Here $C_0$ is the stationary carrier amplitude circulating in the arm
cavities, $k_0$ is the carrier wavenumber, and

\begin{equation}
\mathbf Y_0
=
\left(
1,0,0,\ldots
\right)^T
\label{eq:app_fundamental_vector}
\end{equation}
is the vector corresponding to the fundamental spatial mode. Since the
common mode is assumed to be perfectly mode matched, only the fundamental
arm-cavity mode has a nonzero stationary carrier amplitude and therefore
directly generates radiation-pressure back action.

Substituting Eq.~\eqref{eq:app_returning_field_with_signal} into the SEC
boundary condition \eqref{eq:app_SEC_boundary} gives

\begin{equation}
\mathbf B_s
=
t_s
\left(
I+r_sS
\right)^{-1}
\left[
\Theta_1\mathbf A_s
+
r_s\mathbf D\,\Delta x
\right].
\label{eq:app_Bs_solution}
\end{equation}

The field leaving the signal extraction cavity is

\begin{equation}
\mathbf A_{1s}
=
U_1
\left(
r_s\Theta_1\mathbf A_s+t_s\mathbf B_{1s}
\right),
\label{eq:app_output_boundary}
\end{equation}
which can be reduced to
\begin{equation}
\mathbf A_{1s}
=
W\mathbf A_s-\mathbf P\,\Delta x,
\label{eq:app_output_field}
\end{equation}
where
\begin{align}
W
&=
U_1
\left(
I+r_sS
\right)^{-1}
\left(
r_sI+S
\right)
\Theta_1,
\label{eq:app_W}
\\
\mathbf P
&=
t_sU_1
\left(
I+r_sS
\right)^{-1}
\mathbf D.
\label{eq:app_P}
\end{align}

The matrix $W$ describes the propagation of quantum fluctuations through
the optical system without radiation-pressure back action. The vector
$\mathbf P$ describes the transfer of the differential mirror displacement to
the output optical field.

\subsection{Radiation-pressure back action}
\label{app:radiation_pressure}

The differential displacement satisfies

\begin{equation}
-m\Omega^2\Delta x(\Omega)
=
-mL\Omega^2h(\Omega)
+
F_{\mathrm{BA}}(\Omega).
\label{eq:app_mechanical_equation}
\end{equation}

The radiation-pressure force is determined by the amplitude fluctuations
of the fundamental arm-cavity mode. The effective multimode transfer
matrix from the input field to the arm-cavity field is

\begin{equation}
G(\Omega)
=
M_L
\left(
I-r_{\mathrm{ITM}}M_L^2
\right)^{-1}
\Theta_2M_lM_{L_s}
\left(
I+r_sS
\right)^{-1}
\Theta_1,
\label{eq:app_G_matrix}
\end{equation}

where

\begin{equation}
M_L
=
\left\{
\delta_{nm}
\exp(-ikL)
\exp\left[i(n+1)\Delta\psi_L\right]
\right\}.
\label{eq:app_arm_propagation}
\end{equation}

The matrix elements $g_{0p}(\Omega)$ of the first row of $G$ describe how
quantum fluctuations injected in the $p$-th spatial mode contribute to
the radiation-pressure force acting on the fundamental arm-cavity mode.

The solution of Eq.~\eqref{eq:app_mechanical_equation} is

\begin{align}
\Delta x(\Omega)
&=
Lh(\Omega)
-
\frac{
8\hbar k_0C_0t_{\mathrm{ITM}}t_s
}{
m\Omega^2
}
\sum_p
\Bigg[
\frac{
g_{0p}(\Omega)+g_{0p}^{*}(-\Omega)
}{
2
}
a^a_{sp}(\Omega)
\nonumber\\
&\hspace{4.5cm}
+
i
\frac{
g_{0p}(\Omega)-g_{0p}^{*}(-\Omega)
}{
2
}
a^\phi_{sp}(\Omega)
\Bigg].
\label{eq:app_delta_x_solution}
\end{align}

This expression shows explicitly that spatial-mode mismatch modifies the
radiation-pressure back action by coupling quantum fluctuations from
several input spatial modes to the fundamental arm-cavity mode.

\subsection{Two-photon quadrature representation}
\label{app:two_photon}

For each spatial mode, we introduce the amplitude and phase quadratures

\begin{align}
a^a_{sn}(\Omega)
&=
\frac{
a_{sn}(\Omega)+a^\dagger_{sn}(-\Omega)
}{
\sqrt{2}
},
\label{eq:app_amplitude_quadrature}
\\
a^\phi_{sn}(\Omega)
&=
\frac{
a_{sn}(\Omega)-a^\dagger_{sn}(-\Omega)
}{
i\sqrt{2}
}.
\label{eq:app_phase_quadrature}
\end{align}

The complete input quadrature vector is

\begin{equation}
\mathbf Q_s(\Omega)
=
\left(
a^a_{s0},
a^\phi_{s0},
a^a_{s1},
a^\phi_{s1},
\ldots
\right)^T.
\label{eq:app_input_quadratures}
\end{equation}

The optical transfer matrix $W$ is converted into its two-photon form
$W_q$ by defining

\begin{align}
Y_{nm}^{+}(\Omega)
&=
\frac{
w_{nm}(\Omega)+w_{nm}^{*}(-\Omega)
}{
2
},
\label{eq:app_Y_plus}
\\
Y_{nm}^{-}(\Omega)
&=
\frac{
w_{nm}(\Omega)-w_{nm}^{*}(-\Omega)
}{
2i
}.
\label{eq:app_Y_minus}
\end{align}
Here the coefficients $w_{nm}(\Omega)$ are the elements of $W$.

The resulting quadrature transfer matrix is

\begin{equation}
W_q
=
\begin{pmatrix}
Y_{00}^{+} & -Y_{00}^{-} & Y_{01}^{+} & -Y_{01}^{-} & \cdots \\
Y_{00}^{-} &  Y_{00}^{+} & Y_{01}^{-} &  Y_{01}^{+} & \cdots \\
Y_{10}^{+} & -Y_{10}^{-} & Y_{11}^{+} & -Y_{11}^{-} & \cdots \\
Y_{10}^{-} &  Y_{10}^{+} & Y_{11}^{-} &  Y_{11}^{+} & \cdots \\
\vdots & \vdots & \vdots & \vdots & \ddots
\end{pmatrix}.
\label{eq:app_Wq}
\end{equation}

The displacement-to-output coupling vector in the quadrature basis is

\begin{equation}
\mathbf P_q(\Omega)
=
\begin{pmatrix}
\dfrac{
p_0(\Omega)+p_0^*(-\Omega)
}{
\sqrt{2}
}
\\[1.0em]
\dfrac{
p_0(\Omega)-p_0^*(-\Omega)
}{
i\sqrt{2}
}
\\[1.0em]
\dfrac{
p_1(\Omega)+p_1^*(-\Omega)
}{
\sqrt{2}
}
\\[1.0em]
\dfrac{
p_1(\Omega)-p_1^*(-\Omega)
}{
i\sqrt{2}
}
\\
\vdots
\end{pmatrix}.
\label{eq:app_Pq}
\end{equation}
Here the coefficients $p_{n}(\Omega)$ are the elements of $\mathbf P$.

Similarly, the radiation-pressure coupling vector is

\begin{equation}
\mathbf G_q(\Omega)
=
\frac{
8\hbar k_0C_0t_{\mathrm{ITM}}t_s
}{
m\Omega^2
}
\begin{pmatrix}
\dfrac{
g_{00}(\Omega)+g_{00}^{*}(-\Omega)
}{
2
}
\\[1.0em]
\dfrac{
i\left[
g_{00}(\Omega)-g_{00}^{*}(-\Omega)
\right]
}{
2
}
\\[1.0em]
\dfrac{
g_{01}(\Omega)+g_{01}^{*}(-\Omega)
}{
2
}
\\[1.0em]
\dfrac{
i\left[
g_{01}(\Omega)-g_{01}^{*}(-\Omega)
\right]
}{
2
}
\\
\vdots
\end{pmatrix}.
\label{eq:app_Gq}
\end{equation}

Equation~\eqref{eq:app_delta_x_solution} can therefore be written as

\begin{equation}
\Delta x(\Omega)
=
Lh(\Omega)-\mathbf G_q^T(\Omega)\mathbf Q_s(\Omega).
\label{eq:app_delta_x_quadrature}
\end{equation}

Substitution into Eq.~\eqref{eq:app_output_field} yields

\begin{align}
\mathbf Q_{1s}(\Omega)
&=
W_q(\Omega)\mathbf Q_s(\Omega)
+
P_q(\Omega)\mathbf G_q^T(\Omega)\mathbf Q_s(\Omega)
-
Lh(\Omega)\mathbf P_q(\Omega)
\nonumber\\
&=
Z(\Omega) \mathbf Q_s(\Omega)
-
Lh(\Omega)\mathbf P_q(\Omega),
\label{eq:app_full_IO}
\end{align}
where
\begin{equation}
Z(\Omega)
=
W_q(\Omega)
+
\mathbf P_q(\Omega)\mathbf G_q^T(\Omega).
\label{eq:app_Z}
\end{equation}

\subsection{Output spectral density matrix}
\label{app:spectral_matrix}

Let the input quadrature spectral density matrix be defined by

\begin{equation}
S_0(\Omega)
=
\left\langle
\mathbf Q_s(\Omega) \mathbf Q_s^\dagger(\Omega)
\right\rangle.
\label{eq:app_S0_general}
\end{equation}
Here, $\langle \dots \rangle$ denotes ensemble averaging.

The output spectral density matrix is then

\begin{equation}
S_1(\Omega)
=
Z(\Omega)
S_0(\Omega)
Z^\dagger(\Omega).
\label{eq:app_S1}
\end{equation}

For a phase-squeezed vacuum state occupying only the fundamental spatial
mode, while all higher-order modes are in vacuum states,

\begin{equation}
S_0
=
\operatorname{diag}
\left(
e^{2r},
e^{-2r},
1,
1,
\ldots
\right).
\label{eq:app_squeezed_S0}
\end{equation}

Thus, $S_1(\Omega)$ contains both the spectral densities of individual
output quadratures and the cross-spectral densities between different
spatial modes. These quantities are used in the main text to calculate
the quantum-noise-limited sensitivity, the squeezed and anti-squeezed
quadratures, the squeezing-angle rotation, and the multimode covariance
matrix.

\bibliographystyle{ieeetr}
\bibliography{Misalign.bib, OptoMech.bib}
\end{document}